# *"Shaking" Photons out of a Topological Material*

Mário G. Silveirinha[*]

[(1)] *University of Lisbon–Instituto Superior Técnico and Instituto de Telecomunicações,*

*Avenida Rovisco Pais, 1, 1049-001 Lisboa, Portugal,*

mario.silveirinha@tecnico.ulisboa.pt

**Abstract**

Over the past decade, there has been a great interest in topological effects, with concepts originally developed in the context of electron transport in condensed matter platforms now being extended to optical systems. While topological properties in electronic systems are often linked to the quantization of electric conductivity observed in the integer quantum Hall effect, a direct analogue in optics remains elusive. In this study, we bridge this gap by demonstrating that the response of the Poynting vector (which may be regarded as a "photon current") to the mechanical acceleration of a medium provides a precise photonic analogue of the electric conductivity. In particular, it is shown that the photonic conductivity determines the energy irreversibly transferred from a periodic mechanical driving of the medium to the electromagnetic field. Furthermore, it is demonstrated that for nonreciprocal systems enclosed in a cavity, the constant acceleration of the system induces a flow of photons along a direction perpendicular to the acceleration, analogous to the Hall effect but for light. The spectral density of the photonic conductivity is quantized in the band gaps of the bulk region with the conductivity quantum determined by the gap Chern number.

---

[*] To whom correspondence should be addressed: E-mail: *mario.silveirinha@tecnico.ulisboa.pt*

# I. Introduction

Many similarities and connections between optics and condensed matter physics have been uncovered and studied throughout the years, with recent examples including the emerging field of topological physics [1-9]. An exciting property of topological materials is that they may support topologically non-trivial unidirectional edge states immune to back-scattering [3-5]. In condensed matter physics the nontrivial topology is often associated with the quantization of the electrical conductivity of the relevant materials [10-13], which characterizes the electron transport by an external electromagnetic field, $\mathbf{j} = \boldsymbol{\sigma} \cdot \mathbf{E}$. However, in photonics there is no analogue of the electrical conductivity $\boldsymbol{\sigma}$.

The aim of this article is to bridge the existing gap and demonstrate the existence of a photonic counterpart for $\boldsymbol{\sigma}$. To achieve this goal, we introduce the concept of "photonic conductivity" as the response of the Poynting vector – which can be considered as the "photon current" density – to the mechanical acceleration of a medium. We prove that the photonic conductivity can be written in terms of the Green's function of the relevant material (in the weak dissipation limit, it can be written in terms of the electromagnetic modes of the system), and as an example we compute the photonic conductivity of a Drude plasma with the initial system state ruled by Bose-Einstein statistics (thermal-light). Interestingly, it is found that a photonic conductivity with a nonzero real part implies dissipation and the irreversible transfer of energy from a periodic-in-time mechanical driving into the electromagnetic field. It is relevant to mention that the radiation by moving mirrors [14-20] and related optomechanical systems [21, 22] has been extensively discussed in the literature, e.g., in the context of the dynamical Casimir and Unruh effects [23-30] and sonoluminescence [31]. In this study, we take a novel approach by deriving a linear response function that links the momentum of the thermally generated light and the mechanical acceleration of a medium. Furthermore, we highlight the similarities of this problem with the transport of electrons in solids and the quantum Hall effect.

For the case of a constant acceleration and for a weakly dissipative medium confined inside a cavity, it is shown that the photonic conductivity is determined by an anti-symmetric real-valued tensor, which only depends on the spectral density of the angular momentum of the thermal light in the cavity without acceleration. In particular, for nonreciprocal platforms our theory predicts that the mechanical acceleration of the cavity originates a transverse flow of electromagnetic energy, which is analogous to the Hall-effect observed in condensed-matter systems. Additionally, we show that the photonic conductivity of 2D-type systems with a bulk band gap is precisely quantized, with a conductivity quantum given by the photonic Chern number. Thus, our results demonstrate a direct analogue of the integer quantum Hall effect in photonics [10-13]. It should be noted that topological effects in other types of driven systems were discussed previously by other authors [32-33].

Even though in our theory the excitation is associated with a mechanical driving, most of the developed concepts can be readily extended to the case of spacetime modulated systems [34], where the driving is purely electric. In fact, in recent years it was proven that spacetime modulations of the permittivity and permeability may be used to mimic physical motion [35-46] and induce Doppler type effects [46], synthetic Fresnel drags [39, 40], synthetic Goos-Hanchen shifts [45], synthetic magnetic fields (angular momentum bias) [47-49] and nontrivial topologies [50]. In particular, it was recently shown that a particular subclass of (Minkowskian) spacetime modulations can mimic exactly physical motion [45, 46]. Such systems are formed by isorefractive materials [46] and are the most promising platforms to replicate accelerated physical motions by electronic means. It should be noted that time-varying systems have been amply discussed in the context of the dynamical Casimir effect [27-30].

## II. Photon transport in an accelerated system

### A. Geometry and definitions

Figure 1 depicts a representative geometry of the system studied in this article. It consists of a cavity (box) filled with a possibly nonuniform and dispersive medium. Due to a mechanical driving, the cavity is subject to a shaking-type mechanical motion with frequency $\Omega$. The cavity and all the material structures inside it are modelled as a rigid body. Furthermore, we assume that the radiation inside the cavity is generated by thermal (or quantum) fluctuations.

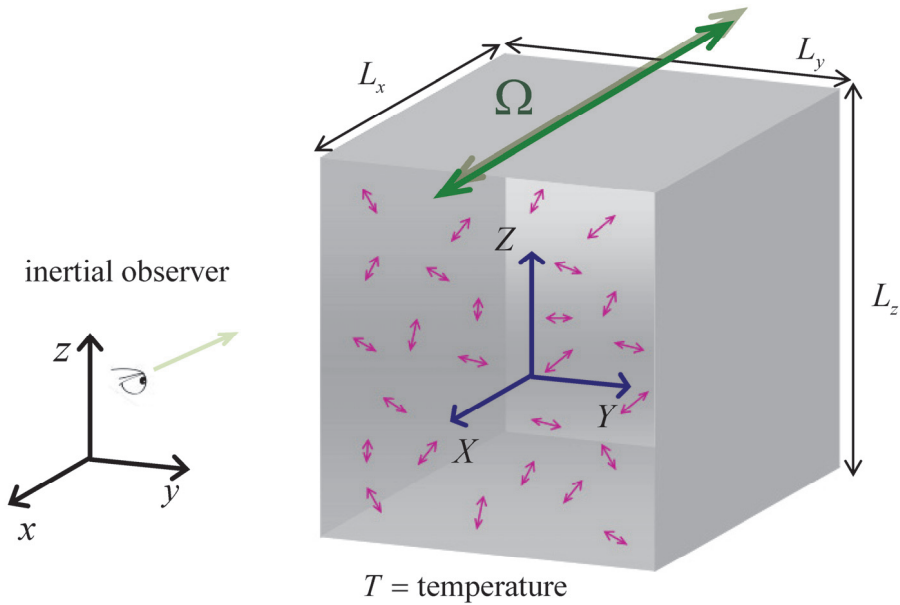

**Fig. 1** A closed box filled with an arbitrary material structure is subject to mechanical shaking with frequency $\Omega$. The light inside the box is originated by thermal and/or quantum fluctuations. The purple arrows represent the equivalent noise sources that generate the thermal light. The reference system $(x,y,z)$ is attached to an inertial (laboratory) frame, whereas the reference system $(X,Y,Z)$ is rigidly attached to the accelerated cavity.

The accelerated motion of the cavity can lead to the emission of light quanta. This phenomenon has been widely studied in the context of the dynamical Casimir effect [23-30]. In the dynamical Casimir effect, the emitted radiation is exclusively due to the accelerated motion of the cavity walls, as the interior of the cavity is typically unfilled (electromagnetic vacuum). In contrast, in our case the interfaces of the material structures inside the box can also lead to the emission of bulk radiation. Furthermore, we shall show that even if the material inside the box is uniform, it is still possible to have bulk radiation provided the medium has a dispersive response.

We use the Poynting vector to characterize the flow of thermal light inside the cavity. The Poynting vector $\mathbf{S}$ may be regarded as the photonic counterpart of the electric current density $\mathbf{j}$, i.e., it may be regarded as the "photon current" density. The analogy is particularly striking for lossless systems (with no light sources in the region of interest), as in such a case $\mathbf{S}$ is linked to the electromagnetic energy density $W$ as $\nabla \cdot \mathbf{S} + \partial_t W = 0$, which the counterpart of the charge continuity equation $\nabla \cdot \mathbf{j} + \partial_t \rho = 0$ (here, $\rho$ is the electric charge density).

In the electronic case, the conductivity gives the response of the electric current (matter) to the applied electric field. Motivated by this definition, we introduce the photonic conductivity ($\boldsymbol{\sigma}_{ph}(\Omega)$) as the response of the "photon current" $\mathbf{S}$ to the acceleration of the medium. Specifically, for an instantaneous acceleration of the form $\mathbf{a}(t) = \mathbf{a}_0 e^{-i\Omega t} + c.c.$ (here c.c. stands for complex conjugation and $\mathbf{a}_0$ is a constant complex-valued vector), the conductivity $\boldsymbol{\sigma}_{ph}(\Omega)$ is defined in such way that:

$$\langle \mathbf{S}_{av}(t) \rangle = -\frac{1}{c^2} \boldsymbol{\sigma}_{ph}(\Omega) \cdot \mathbf{a}_0 e^{-i\Omega t} + c.c.. \tag{1}$$

The conductivity $\boldsymbol{\sigma}_{ph}$ depends on the oscillation frequency $\Omega$ and has unities of W/m. In the above, $\langle \mathbf{S}_{av}(t) \rangle$ is the thermal expectation value of the Poynting vector averaged over the cavity volume:

$$\langle \mathbf{S}_{av} \rangle = \frac{1}{V} \int_{box} \langle \mathbf{S} \rangle d^3\mathbf{r}. \tag{2}$$

The symbol $\langle ... \rangle$ represents the statistical expectation and the symbol "av" refers to the volume averaging.

The (Abraham) electromagnetic momentum density, i.e., the light momentum per unit of volume, is linked to the Poynting vector as $\mathbf{g} = \mathbf{S}/c^2$ [51-55]. Therefore, $\langle \mathbf{g}_{av} \rangle \equiv \langle \mathbf{S}_{av} \rangle / c^2$ is precisely the expectation of the light momentum inside the box divided by the volume. In particular, the photonic conductivity can also be regarded as the response of the expectation of the (Abraham) light momentum to the mechanical acceleration:

$$\langle \mathbf{g}_{av}(t) \rangle = -\frac{1}{c^4} \boldsymbol{\sigma}_{ph}(\Omega) \cdot \mathbf{a}_0 e^{-i\Omega t} + c.c.. \tag{3}$$

It is worth noting that for a sufficiently large cavity the photonic conductivity may not be influenced by the material properties of the cavity walls. This is because the light emission from the bulk region tends to dominate, making the photonic conductivity a bulk medium property. However, there may be instances where the contributions from the boundary walls become significant, particularly when the cavity is unfilled and there is no radiation from the bulk region. In these cases, the boundary walls can play an important role in determining the photonic conductivity. Additionally, we will demonstrate in Sect. V that in topological systems, the cavity walls play an essential role.

## B. Coordinate transformations

In order to study the interaction of the electromagnetic field with accelerated material bodies it is convenient to switch to a set of coordinates $\mathbf{R} = (X, Y, Z)$ and $\tau = t$ rigidly attached to the cavity. Loosely speaking, we will refer to $\mathbf{R}, \tau$ as the co-moving frame coordinates, even though they are not associated with an inertial reference frame. The coordinates in the inertial laboratory frame are $\mathbf{r}, t$ with $\mathbf{r} = (x, y, z)$. The link between the two sets of spatial coordinates is $\mathbf{r} = \mathbf{R} + \mathbf{r}_0(\tau)$ with $\mathbf{r}_0(\tau)$ the coordinates of a generic point of the cavity (the "origin") as a function of time. The trajectory $\mathbf{r}_0(\tau)$ is completely determined by the mechanical driving. The instantaneous velocity and acceleration are given by the first and second order derivatives in time of $\mathbf{r}_0(\tau)$, respectively.

We demonstrate in Appendix A that the electrodynamics of the accelerated cavity can be conveniently studied by introducing a set of fields $\mathbf{f}_{co} = \begin{pmatrix} \mathbf{E}_{co} & \mathbf{H}_{co} \end{pmatrix}^T$, which can be written in terms of the fields in the laboratory frame as in Eq. (A2). We shall loosely refer to $\mathbf{f}_{co} = \mathbf{f}_{co}(\mathbf{R}, \tau)$ as the co-moving frame fields. In the non-relativistic limit, the time evolution of $\mathbf{f}_{co}$ is determined by a differential system, which can be written in a compact manner as:

$$\hat{L} \cdot \mathbf{f}_{co} + \hat{L}_{int} \cdot \mathbf{f}_{co} = i \frac{\partial}{\partial \tau} \mathbf{g}_{co} + i \mathbf{j}_{ext,co} \ . \tag{4}$$

In the above, $\hat{L} = \hat{L}(-i\nabla_{\mathbf{R}})$ is a differential operator determined by the curl operators of the Maxwell's equations [Eq. (A10)]. Furthermore, $\mathbf{g}_{co} = \begin{pmatrix} \mathbf{D}_{co} & \mathbf{B}_{co} \end{pmatrix}^T$ is linked to $\mathbf{f}_{co} = \begin{pmatrix} \mathbf{E}_{co} & \mathbf{H}_{co} \end{pmatrix}^T$ exactly by the same constitutive relations as in a system without the mechanical driving. For simplicity, we assume that all the materials in the cavity are non-magnetic so that $\mathbf{B}_{co} = \mu_0 \mathbf{H}_{co}$. Furthermore, $\mathbf{j}_{ext,co}$ represents the external electromagnetic currents, e.g., thermal noise currents, in the co-moving frame coordinates. The effect of the acceleration is described by the interaction operator $\hat{L}_{int} = -i\frac{\mathbf{a}}{c^2} \cdot \hat{\mathbf{S}} - i\frac{\mathbf{v}}{c^2} \cdot \hat{\mathbf{S}} \partial_\tau$. Here, $\mathbf{v}$ and $\mathbf{a}$ are the instantaneous (time dependent) velocity and acceleration, respectively, $\partial_\tau = \partial/\partial \tau$, and $\hat{\mathbf{S}}$ is a tensor determined by Eq. (A11). In the co-moving frame all the material structures confined within the cavity have time-independent coordinates: the effect of the accelerated motion is fully modeled by $\hat{L}_{int}$. Without the mechanical driving ($\mathbf{v} = 0 = \mathbf{a}$), the master equation (4) reduces to the standard Maxwell's equations in the undriven cavity. The fields in the laboratory frame coordinates can be found from the solution of (4) using $\mathbf{E}(\mathbf{r},t) = \mathbf{E}_{co} - \mathbf{v} \times \mathbf{B}_{co}$ and $\mathbf{B}(\mathbf{r},t) = \mathbf{B}_{co} + \frac{1}{c^2} \mathbf{v} \times \mathbf{E}_{co}$ with $\mathbf{R} = \mathbf{r} - \mathbf{r}_0(\tau)$ and $\tau = t$. The reader is referred to Appendix A for additional discussion.

## C. Electromagnetic inertia

In a cavity at rest, the distribution of thermal-light energy is expected to be relatively uniform. In contrast, when the cavity is subject to an accelerated motion towards the +x direction, the wall $X = -L_x/2$ effectively "closes in" on the thermal-light field, leading to an increased energy density near this wall. At the same time, the opposite wall ($X = +L_x/2$) moves away from the thermal light field, resulting in a depleted energy density near that wall. This property suggests that the acceleration may cause the accumulation of electromagnetic energy to on the wall ($X = -L_x/2$) opposite to the direction of acceleration (+x-direction).

We formally demonstrate this property in Appendix B, by showing that the modes of a cavity subject to a constant acceleration are exactly the same as the modes of the undriven cavity apart from an exponential decay factor $e^{-X\frac{a}{c^2}}$. The bunching of energy on the back cavity wall may be regarded as a consequence of *electromagnetic inertia*. It implies that the thermal-light is unevenly distributed inside the accelerated cavity, tending to pile up on the wall $X = -L_x/2$, akin to a temperature gradient caused by the acceleration.

Evidently, when the box is subject to a time-varying acceleration, $\mathbf{a} = \mathbf{a}_0 e^{-i\Omega t} + c.c.$, the distribution of thermal-light energy in the box must adjust itself dynamically to the mechanical oscillation, leading thereby to an oscillation of the Poynting vector, i.e., to a time-varying "photon" current.

It relevant to point out at this point that our analysis neglects any local temperature variations of the cavity walls arising from frictional effects due to the acceleration of the box in air. It implicitly assumes that all the material structures are rigid and that the noise sources are accelerated with the box.

## III. Photonic conductivity

In this section, we derive an explicit formula for the photonic conductivity $\boldsymbol{\sigma}_{ph}(\Omega)$, showing that it can be written in terms of the system Green's function.

### A. Linear response function

For non-magnetic materials, the Poynting vector $\mathbf{S} = \mathbf{E} \times \mathbf{B}/\mu_0$ in the laboratory frame can be expressed in terms of the fields $\mathbf{f}_{co} = (\mathbf{E}_{co} \quad \mathbf{H}_{co})^T$ as [Eq. (A2)]:

$$\mathbf{S} = (\mathbf{E}_{co} - \mathbf{v} \times \mathbf{B}_{co}) \times \left(\frac{\mathbf{B}_{co}}{\mu_0} + \mathbf{v} \times \varepsilon_0 \mathbf{E}_{co}\right) = \mathbf{S}_{co} + \delta \mathbf{S}_{co} + o(v^2) \tag{5}$$

with,

$$\mathbf{S}_{co} = \mathbf{E}_{co} \times \mathbf{H}_{co}, \qquad \delta \mathbf{S}_{co} = \overline{\mathbf{U}}(\mathbf{f}_{co}) \cdot \mathbf{v}, \tag{6a}$$

$$\overline{\mathbf{U}}(\mathbf{f}_{co}) \equiv -(\varepsilon_0 \mathbf{E}_{co} \otimes \mathbf{E}_{co} + \mu_0 \mathbf{H}_{co} \otimes \mathbf{H}_{co}) + (\varepsilon_0 \mathbf{E}_{co} \cdot \mathbf{E}_{co} + \mu_0 \mathbf{H}_{co} \cdot \mathbf{H}_{co}) \mathbf{1}_{3\times3}. \tag{6b}$$

We used $\mathbf{H}_{co} = \mathbf{B}_{co}/\mu_0$ and ignored terms that are of second order ($o(v^2)$) in the velocity. Note that $\overline{\mathbf{U}}$ may be regarded as the energy-momentum tensor. Substituting the above formulas into Eq. (2) and using the coordinate transformation $\mathbf{R} = \mathbf{r} - \mathbf{r}_0(t)$ and $\tau = t$, one sees that the photon current can be written as:

$$\langle \mathbf{S}_{av}(t) \rangle = \frac{1}{V} \int_{box} \langle \mathbf{S}_{co} + \overline{\mathbf{U}}(\mathbf{f}_{co}) \cdot \mathbf{v} \rangle d^3 \mathbf{R} \tag{7}$$

As previously discussed, we want to find the response of $\langle \mathbf{S}_{av}(t) \rangle$ to the mechanical driving (acceleration) when the light inside the box is generated by thermal fluctuations, i.e., by the "noise" currents $\mathbf{j}_N$. Let us write the fluctuation-induced fields $\mathbf{f}_{co}$ in the co-moving frame coordinates as a sum of the fluctuation-fields without the mechanical driving ($\mathbf{f}_{co}^{0} = \begin{pmatrix} \mathbf{E}_{co}^{0} & \mathbf{H}_{co}^{0} \end{pmatrix}^{T}$) and a perturbation due to the mechanical driving ($\mathbf{f}_{co}^{int} = \begin{pmatrix} \mathbf{E}_{co}^{int} & \mathbf{H}_{co}^{int} \end{pmatrix}^{T}$): $\mathbf{f}_{co} = \mathbf{f}_{co}^{0} + \mathbf{f}_{co}^{int}$. By substituting $\mathbf{f}_{co} = \mathbf{f}_{co}^{0} + \mathbf{f}_{co}^{int}$ into Eq. (7) and retaining only the terms that are linear in the velocity (linear response theory), it is found that:

$$\langle \mathbf{S}_{av}(t) \rangle = \frac{1}{V} \int_{box} \langle \mathbf{E}_{co}^{0} \times \mathbf{H}_{co}^{int} + \mathbf{E}_{co}^{int} \times \mathbf{H}_{co}^{0} \rangle d^{3}\mathbf{R} + \overline{\mathbf{U}}^{0} \cdot \mathbf{v}. \tag{8}$$

with $\overline{\mathbf{U}}^{0} \equiv \frac{1}{V} \int_{box} \langle \overline{\mathbf{U}}(\mathbf{f}_{co}^{0}) \rangle d^{3}\mathbf{R}$ a constant symmetric and real-valued matrix that only depends on the unperturbed fluctuation-fields. The previous formula can also be written as:

$$\langle \mathbf{S}_{av}(t) \rangle = \frac{2}{V} \langle \mathbf{f}_{co}^{0} | \hat{\mathbf{S}} | \mathbf{f}_{co}^{int} \rangle + \overline{\mathbf{U}}^{0} \cdot \mathbf{v}, \tag{9}$$

where $\langle ... | ... \rangle$ is the canonical inner product defined by $\langle \mathbf{f}_{1} | \mathbf{f}_{2} \rangle = \frac{1}{2} \int \mathbf{f}_{1}^{*} \cdot \mathbf{f}_{2} d^{3}\mathbf{R}$ and $\hat{\mathbf{S}}$ is the operator defined in Appendix A (Eq. (A11)). We took into account that the fields are real-valued. Note that $\langle \mathbf{f}_{co} | \hat{\mathbf{S}} | \mathbf{f}_{co} \rangle / V = (\mathbf{E}_{co} \times \mathbf{H}_{co})_{av} \equiv \mathbf{S}_{co,av}$ is the volume averaged Poynting vector in the co-moving frame coordinates.

## B. Noise currents and the thermal-light correlations

In order to find the perturbation due to the mechanical driving, it is convenient to introduce the Green's function $\overline{\mathcal{G}} = \overline{\mathcal{G}}(\mathbf{r}, \mathbf{r}'; \omega)$ of the unperturbed system. It is defined in such a way that:

$$(\hat{L} - \omega \mathbf{M}(\mathbf{r}, \omega)) \cdot \overline{\mathcal{G}} = \omega \mathbf{1} \delta(\mathbf{r} - \mathbf{r}'). \tag{10}$$

Here, $\mathbf{M}(\mathbf{r}, \omega)$ is the material response that links $\mathbf{g}_{co} = \begin{pmatrix} \mathbf{D}_{co} & \mathbf{B}_{co} \end{pmatrix}^T$ and $\mathbf{f}_{co} = \begin{pmatrix} \mathbf{E}_{co} & \mathbf{H}_{co} \end{pmatrix}^T$ in the frequency domain (without the mechanical acceleration): $\mathbf{g}_{co,\omega} = \mathbf{M}(\mathbf{r}, \omega) \cdot \mathbf{f}_{co,\omega}$. For a standard dielectric it is of the form:

$$\mathbf{M}(\mathbf{r}, \omega) = \begin{pmatrix} \varepsilon_0 \overline{\varepsilon} & \mathbf{0}_{3\times 3} \\ \mathbf{0}_{3\times 3} & \mu_0 \mathbf{1}_{3\times 3} \end{pmatrix}, \tag{11}$$

where $\overline{\varepsilon}(\mathbf{r}, \omega)$ is the relative permittivity tensor. The fluctuation dissipation theorem relates the correlations of the noise currents $\mathbf{j}_N$ that create the fluctuation fields with the material loss [56, 57]:

$$\frac{1}{(2\pi)^2} \langle \mathbf{j}_{N,\omega}(\mathbf{r}) \mathbf{j}^*_{N,\omega'}(\mathbf{r}') \rangle = \frac{\omega \mathcal{E}_\omega}{\pi} \frac{1}{2i} \left[ \mathbf{M}(\mathbf{r}, \omega) - \mathbf{M}^\dagger(\mathbf{r}, \omega) \right] \delta(\omega - \omega') \delta(\mathbf{r} - \mathbf{r}'). \tag{12}$$

Here, $\mathcal{E}_\omega = \frac{\hbar\omega}{2} + \frac{\hbar\omega}{\exp(\hbar\omega/k_B T) - 1} = \frac{\hbar\omega}{2} \coth\left(\frac{\hbar\omega}{2k_B T}\right)$ is the energy of a quantum harmonic oscillator at temperature $T$, and the dagger symbol † represents the Hermitian conjugate matrix. In the frequency domain, the unperturbed fluctuation fields $\mathbf{f}^0$ are linked to the noise currents through the Green's function:

$$\mathbf{f}^0_\omega(\mathbf{r}) = \int \overline{\mathcal{G}}(\mathbf{r}, \mathbf{r}'; \omega) \cdot \frac{1}{-i\omega} \mathbf{j}_{N,\omega}(\mathbf{r}') d^3\mathbf{r}'. \tag{13}$$

The correlations of the thermal-light fields are determined by the Green's function [56, 57]:

$$\frac{1}{(2\pi)^2} \langle \mathbf{f}^0_\omega(\mathbf{r}) \mathbf{f}^{0*}_{\omega'}(\mathbf{r}') \rangle = \frac{\mathcal{E}_\omega}{\pi \omega} \delta(\omega - \omega') \frac{1}{2i} \left[ \overline{\mathcal{G}}(\mathbf{r}, \mathbf{r}', \omega) - \overline{\mathcal{G}}^\dagger(\mathbf{r}', \mathbf{r}, \omega) \right]. \tag{14}$$

### C. Perturbation due to the mechanical driving

With a mechanical driving, the fluctuation fields $\mathbf{f}_{co}$ satisfy Eq. (4) with the excitation in the instantaneously co-moving frame determined by the noise currents $\mathbf{j}_{ext,co} = \mathbf{j}_N$. We can obtain a perturbative (series) solution in terms of the strength of the interaction ($\hat{L}_{int}$). Evidently, the zero-order term is simply $\mathbf{f}_{co}^0 = \mathbf{f}^0$ with $\mathbf{f}^0$ determined by Eq. (13), so that in the time-domain:

$$\mathbf{f}_{co}^0(\mathbf{r},t) = \frac{1}{2\pi} \int_{-\infty}^{+\infty} d\omega \, \mathbf{f}_{\omega}^0(\mathbf{r}) e^{-i\omega t}. \tag{15}$$

For notational simplicity, we solve the differential system (4) with $\mathbf{R} \to \mathbf{r}$ and $\tau \to t$, but it should be clear that all the calculations are done in the co-moving frame coordinates. Feeding the zero-order solution back to Eq. (4), one sees that it induces the additional current $i\hat{L}_{int} \cdot \mathbf{f}_{co}^0$. Hence, the first-order correction to the unperturbed solution is:

$$\mathbf{f}_{co}^{int}(\mathbf{r},t) = \frac{1}{2\pi} \int_{-\infty}^{+\infty} d\omega \, e^{-i\omega t} \int d^3\mathbf{r}' \, \overline{\mathcal{G}}(\mathbf{r},\mathbf{r}';\omega) \cdot \frac{-1}{\omega} \left[\hat{L}_{int} \cdot \mathbf{f}_{co}^0\right]_\omega. \tag{16}$$

The term in rectangular brackets represents the Fourier transform in time of $\hat{L}_{int} \cdot \mathbf{f}_{co}^0$. To proceed, we assume that the velocity oscillates in time with frequency $\Omega$, so that $\mathbf{v} = \mathbf{v}_0 e^{-i\Omega t} + c.c.$ and $\mathbf{a} = \mathbf{a}_0 e^{-i\Omega t} + c.c.$ with $\mathbf{a}_0 = -i\Omega \mathbf{v}_0$. Then, using $\hat{L}_{int} = -i\frac{\mathbf{a}}{c^2} \cdot \hat{\mathbf{S}} - i\frac{\mathbf{v}}{c^2} \cdot \hat{\mathbf{S}} \partial_\tau$ in Eq. (15), it is readily found that:

$$\left[\hat{L}_{int} \cdot \mathbf{f}_{co}^0\right]_\omega = -\left[\frac{i\mathbf{a}_0}{c^2} \cdot \hat{\mathbf{S}} + (\omega - \Omega)\frac{\mathbf{v}_0}{c^2} \cdot \hat{\mathbf{S}}\right] \cdot \mathbf{f}_{\omega-\Omega}^0(\mathbf{r})$$
$$- \left[\frac{i\mathbf{a}_0^*}{c^2} \cdot \hat{\mathbf{S}} + (\omega + \Omega)\frac{\mathbf{v}_0^*}{c^2} \cdot \hat{\mathbf{S}}\right] \cdot \mathbf{f}_{\omega+\Omega}^0(\mathbf{r}) \tag{17}$$

Thereby, it follows that the fields induced by the mechanical driving are given by:

$$\mathbf{f}_{co}^{int}(\mathbf{r},t) = \frac{1}{2\pi} \int_{-\infty}^{+\infty} d\omega' \int d^3\mathbf{r}' \, \overline{\mathcal{G}}(\mathbf{r},\mathbf{r}',\omega') \cdot \frac{\mathbf{v}_0}{c^2} \cdot \hat{\mathbf{S}} \cdot \mathbf{f}_{\omega'-\Omega}^0(\mathbf{r}') e^{-i\omega' t}$$
$$+ \frac{1}{2\pi} \int_{-\infty}^{+\infty} d\omega' \int d^3\mathbf{r}' \, \overline{\mathcal{G}}(\mathbf{r},\mathbf{r}',\omega') \cdot \frac{\mathbf{v}_0^*}{c^2} \cdot \hat{\mathbf{S}} \cdot \mathbf{f}_{\omega'+\Omega}^0(\mathbf{r}') e^{-i\omega' t} \tag{18}$$

## D. Derivation of the photonic conductivity

With the help of Eqs. (15) and (18), we can evaluate the term $\langle \mathbf{f}_{co}^0 | \hat{\mathbf{S}} | \mathbf{f}_{co}^{int} \rangle$ in Eq. (9). It is given by:

$$2\langle \mathbf{f}_{co}^0 | \hat{\mathbf{S}} | \mathbf{f}_{co}^{int} \rangle = \frac{1}{(2\pi)^2} \int_{-\infty}^{+\infty} d\omega \int_{-\infty}^{+\infty} d\omega' \int d^3\mathbf{r} \int d^3\mathbf{r}' \mathbf{f}_\omega^{0,*}(\mathbf{r}) \cdot \hat{\mathbf{S}} \cdot \overline{\mathcal{G}}(\mathbf{r},\mathbf{r}',\omega') \cdot \frac{\mathbf{v}_0}{c^2} \cdot \hat{\mathbf{S}} \cdot \mathbf{f}_{\omega'-\Omega}^0(\mathbf{r}') e^{-i(\omega'-\omega)t}$$
$$+ \frac{1}{(2\pi)^2} \int_{-\infty}^{+\infty} d\omega \int_{-\infty}^{+\infty} d\omega' \int d^3\mathbf{r} \int d^3\mathbf{r}' \mathbf{f}_\omega^{0,*}(\mathbf{r}) \cdot \hat{\mathbf{S}} \cdot \overline{\mathcal{G}}(\mathbf{r},\mathbf{r}',\omega') \cdot \frac{\mathbf{v}_0^*}{c^2} \cdot \hat{\mathbf{S}} \cdot \mathbf{f}_{\omega'+\Omega}^0(\mathbf{r}') e^{-i(\omega'-\omega)t}$$

(19)

Using the fluctuation dissipation theorem [Eq. (14)], we take the statistical expectation of the right-hand side. After some analysis, one finds that:

$$2\langle \mathbf{f}_{co}^0 | \hat{\mathbf{S}} | \mathbf{f}_{co}^{int} \rangle = \mathbf{\Sigma}(\Omega) \cdot \mathbf{v}_0 e^{-i\Omega t} + c.c. \tag{20}$$

where $\mathbf{\Sigma}(\Omega)$ is the tensor with generic components $i$ and $j$:

$$\Sigma_{ij}(\Omega) = \frac{1}{c^2} \int_{-\infty}^{+\infty} d\omega \mathcal{E}_\omega \int d^3\mathbf{r} \int d^3\mathbf{r}' \mathrm{Tr}\left\{ \hat{\mathbf{S}}_i \cdot \overline{\mathcal{G}}(\mathbf{r},\mathbf{r}',\omega+\Omega) \cdot \hat{\mathbf{S}}_j \cdot \frac{1}{2\pi i \omega}\left[ \overline{\mathcal{G}}(\mathbf{r}',\mathbf{r},\omega) - \overline{\mathcal{G}}^\dagger(\mathbf{r},\mathbf{r}',\omega) \right] \right\}$$

(21)

In the above, $\hat{\mathbf{S}}_i$ are the matrices defined in Eq. (A11) and $\mathrm{Tr}\{...\}$ represents the trace of a matrix. We used the reality property $\overline{\mathcal{G}}(\mathbf{r},\mathbf{r}',\omega) = \overline{\mathcal{G}}^*(\mathbf{r},\mathbf{r}',-\omega^*)$. Combining the above result with Eq. (9), we conclude that the (linear) response of the photon current to the mechanical driving ($\mathbf{v} = \mathbf{v}_0 e^{-i\Omega t} + c.c.$) is such that:

$$\langle \mathbf{S}_{av}(t) \rangle = \left[ \frac{1}{V} \mathbf{\Sigma}(\Omega) + \overline{\mathbf{U}}^0 \right] \cdot \mathbf{v}_0 e^{-i\Omega t} + c.c.. \tag{22}$$

The response function is the term in rectangular brackets. In principle, it is possible to have a nontrivial response even for a linear motion i.e., when $\Omega = 0$. In fact, for $\mathbf{v} = const.$ an observer in the laboratory frame sees the thermal-light energy in the box travelling with a constant velocity, and this may correspond to a nontrivial flow. Evidently, such a "convective-type" term is not associated with light-matter interactions and hence is not relevant for our purposes. We aim to characterize the perturbation of the Poynting vector expectation due to the *acceleration* as defined by Eq. (1). Clearly, from the previous considerations, it is given by $\boldsymbol{\sigma}_{ph}(\Omega) = \frac{1}{i\Omega} \frac{c^2}{V} [\boldsymbol{\Sigma}(\Omega) - \boldsymbol{\Sigma}(0)]$ where we eliminated the part of the response ($\frac{1}{V}\boldsymbol{\Sigma}(0) + \overline{\mathbf{U}}^0$) that is responsible for the convective flow in the case of a linear uniform motion with $\mathbf{v} = const.$. The photonic conductivity tensor can be explicitly written as:

$$\boldsymbol{\sigma}_{ph,ij}(\Omega) = \frac{-1}{V} \int_{-\infty}^{+\infty} d\omega \, \mathcal{E}_\omega \int d^3\mathbf{r} \int d^3\mathbf{r}'$$
$$\text{Tr}\left\{ \hat{\mathbf{S}}_i \cdot \frac{1}{\Omega} \left[ \overline{\mathcal{G}}(\mathbf{r},\mathbf{r}',\omega+\Omega) - \overline{\mathcal{G}}(\mathbf{r},\mathbf{r}',\omega) \right] \cdot \hat{\mathbf{S}}_j \cdot \frac{1}{2\pi\omega} \left[ \overline{\mathcal{G}}(\mathbf{r}',\mathbf{r},\omega) - \overline{\mathcal{G}}^\dagger(\mathbf{r},\mathbf{r}',\omega) \right] \right\}$$
(23)

When the undriven system is reciprocal the photonic conductivity tensor is transpose symmetric, consistent with the Onsager principle [58]. The proof relies on Eq. (E9) and is omitted for conciseness. In Appendix C, we obtain a modal expansion for the photonic conductivity applicable in the weak dissipation limit.

It is convenient to introduce a (unilateral) spectral density of the photonic conductivity ($\boldsymbol{\sigma}_{ph}(\Omega) = \int_{0^+}^{+\infty} d\omega \, \boldsymbol{\sigma}_{ph,\omega}(\Omega)$) determined by:

$$\boldsymbol{\sigma}_{ph,\omega,ij}(\Omega) = \boldsymbol{\sigma}^b_{ph,\omega,ij}(\Omega) + \boldsymbol{\sigma}^b_{ph,-\omega,ij}(\Omega),$$
(24a)

$$\boldsymbol{\sigma}^b_{ph,\omega,ij}(\Omega) = \frac{-\mathcal{E}_\omega}{V} \int d^3\mathbf{r} \int d^3\mathbf{r}'$$
$$\text{Tr}\left\{ \hat{\mathbf{S}}_i \cdot \frac{1}{\Omega} \left[ \overline{\mathcal{G}}(\mathbf{r},\mathbf{r}',\omega+\Omega) - \overline{\mathcal{G}}(\mathbf{r},\mathbf{r}',\omega) \right] \cdot \hat{\mathbf{S}}_j \cdot \frac{1}{2\pi} \left[ \frac{\overline{\mathcal{G}}(\mathbf{r}',\mathbf{r},\omega)}{\omega} - \left(\frac{\overline{\mathcal{G}}(\mathbf{r},\mathbf{r}',\omega)}{\omega}\right)^\dagger \right] \right\}$$
(24b)

It is simple to check that $\boldsymbol{\sigma}^b_{\text{ph},-\omega,ij}(\Omega) = \left[\boldsymbol{\sigma}^b_{\text{ph},\omega^*,ij}(-\Omega)\right]^*$. Note that $\boldsymbol{\sigma}^b_{\text{ph},\omega,ij}(\Omega)$ is the bi-lateral spectral density defined for positive and negative $\omega$. The spectral density $\boldsymbol{\sigma}_{\text{ph},\omega}(\Omega)$ determines the part of the "photon current" associated with radiation with frequency $\omega$ that is induced by the mechanical acceleration, $\langle \mathbf{S}_{\text{av},\omega}(t) \rangle = -\frac{1}{c^2} \boldsymbol{\sigma}_{\text{ph},\omega}(\Omega) \cdot \mathbf{a}_0 e^{-i\Omega t} + c.c.$.

## IV. Irreversible Light Emission

### A. Power transferred from the mechanical driving to the electromagnetic field

As discussed in Sect. II.C, the acceleration of the box leads to an asymmetric distribution of the thermal light energy within the box, and to an oscillating light flow. It is natural to wonder if in a full oscillating cycle there is net transfer of energy from the mechanical driving to the electromagnetic field. In order to investigate this aspect, we note that from the stress-tensor theorem the Lorentz force ($\mathbf{F}_L$) acting on a set of particles within a volume $V$ is related to the electromagnetic stress tensor $\overline{\mathbf{T}}$ as follows [55]:

$$\int_{\partial V} ds\, \overline{\mathbf{T}} \cdot \hat{\mathbf{n}} = \mathbf{F}_L + \frac{d}{dt} \int_V d^3\mathbf{r}\, \frac{1}{c^2} \mathbf{S}. \qquad (25)$$

The integral on the left gives the flux of the stress-tensor over the boundary surface of $V$. If this surface encloses completely the cavity walls, it only depends on the fields external to the box, and hence it may be identified with the "mechanical" force ($\mathbf{F}_{\text{drive}}$). On the other hand, from Newton's law, the Lorentz force equals $M\mathbf{a}$, with $M$ the total mass of the box. Finally, the integral $\int_V d^3\mathbf{r}\, \frac{1}{c^2}\mathbf{S}$ gives the light momentum inside the box, whose expectation is linked to the acceleration through Eq. (3). Thus, the previous discussion shows that:

$$\mathbf{F}_{\text{drive}} = M\mathbf{a} + V\frac{d}{dt}\langle \mathbf{g}_{\text{av}}(t)\rangle. \tag{26}$$

It is interesting to note that the term $\frac{d}{dt}\langle \mathbf{g}_{\text{av}}(t)\rangle$ may be regarded as a radiation-reaction term, as it is proportional to the derivative of the acceleration, analogous to the Abraham-Lorentz self-force for a point charge [59]. The instantaneous power transferred from the mechanical driving to the electromagnetic field ($p_{d\to\text{EM}}(t)$) is given by:

$$\frac{p_{d\to\text{EM}}(t)}{V} = \mathbf{v}(t)\cdot \frac{d}{dt}\langle \mathbf{g}_{\text{av}}(t)\rangle, \tag{27}$$

with $\mathbf{v}(t)$ the velocity of the box. For a time-harmonic mechanical oscillation with $\mathbf{a}(t) = \mathbf{a}_0 e^{-i\Omega t} + c.c.$ and $\mathbf{v}(t) = \frac{\mathbf{a}_0}{-i\Omega}e^{-i\Omega t} + c.c.$ the time-averaged power (per unit of volume) is:

$$\frac{\langle p_{d\to\text{EM}}\rangle}{V} = \frac{2}{c^4}\text{Re}\{\mathbf{a}_0^* \cdot \boldsymbol{\sigma}_{\text{ph}}(\Omega)\cdot \mathbf{a}_0\}. \tag{28}$$

Hence, the Hermitian part of the conductivity tensor $\frac{\boldsymbol{\sigma}_{\text{ph}} + \boldsymbol{\sigma}_{\text{ph}}^\dagger}{2}$ controls the irreversible light-matter interactions. In principle, $\frac{\boldsymbol{\sigma}_{\text{ph}} + \boldsymbol{\sigma}_{\text{ph}}^\dagger}{2}$ must be positive definite (or nonnegative) to ensure that the energy is transferred from the mechanical driving to the field, and not the other way around. We mention in passing that the term $\frac{1}{V}\boldsymbol{\Sigma}(0) + \overline{\mathbf{U}}^0$ dropped after Eq. (22) would not contribute to $\langle p_{d\to\text{EM}}\rangle$ for any linear mechanical oscillation, had it been retained. Furthermore, it can be shown that Eq. (28) agrees exactly with the result obtained from the intuitive formula (proof is omitted for conciseness):

$$\frac{\langle p_{d \to EM} \rangle}{V} = -\frac{1}{T} \int dt \, \langle \mathbf{f}_{co} \cdot \mathbf{j}_{co} \rangle_{av} . \tag{29}$$

with $\mathbf{f}_{co} = \mathbf{f}_{co}^0 + \mathbf{f}_{co}^{int}$ the total fields and $\mathbf{j}_{co} = \mathbf{j}_N + i\hat{L}_{int} \cdot \mathbf{f}_{co}^0$ the equivalent external current. As before, the $\langle ... \rangle$ represents the statistical expectation and "av" refers to volume averaging. The time integral in the above equation is over one oscillation period ($T = 2\pi / \Omega$). Note that $-\mathbf{f}_{co} \cdot \mathbf{j}_{co}$ is the power drawn from the external currents per unit of volume.

The energy transferred to the field is dissipated in the form of heat (see subsection IV.C). It is useful to briefly comment on the difference between the dissipation mechanisms in photonic and electronic cases. In electronic transport problems, the energy is dissipated through charge collisions with the other charges. In contrast, the classical electromagnetic field does not self-interact and hence the dissipation must be mediated by interactions with matter.

### B. Photonic conductivity for a Drude permittivity dispersion

In order to illustrate the ideas developed so far, let us consider the case where the box is filled uniformly by a non-magnetic material with the Drude permittivity dispersion $\varepsilon / \varepsilon_0 = 1 - \omega_p^2 / \omega(\omega + i\Gamma)$, with $\omega_p$ the plasma frequency and $\Gamma$ the collision frequency. In Appendix D, we prove formally that (for a sufficiently large box) the corresponding photonic conductivity tensor is a scalar $\sigma_{ph}(\Omega)$ independent of the box dimensions.

The irreversible interactions between the external mechanical driving and the medium are described by the real part of the conductivity $\text{Re}\{\sigma_{ph}(\Omega)\}$. It is shown in Appendix D, that in the weak dissipation limit $\Gamma \to 0^+$ and for $\Omega > 0$ the spectral density of the photonic conductivity is [Eq. (D13)]:

$$\text{Re}\{\sigma_{ph,\omega}(\Omega)\} = \frac{\mathcal{E}_\omega}{12\pi c}(\Omega+2\omega_p)\frac{\sqrt{(\Omega+\omega_p)^2-\omega_p^2}}{\Omega+\omega_p}\left[-\omega_p\delta(\omega-(\Omega+\omega_p))+(\Omega+\omega_p)\delta(\omega-\omega_p)\right]$$
$$+\frac{\mathcal{E}_\omega}{12\pi c}(\Omega-2\omega_p)\frac{\sqrt{(\Omega-\omega_p)^2-\omega_p^2}}{\Omega-\omega_p}\left[+\omega_p\delta(\omega-(\Omega-\omega_p))+(\Omega-\omega_p)\delta(\omega-\omega_p)\right]$$

(30)

In the $\omega_p \to 0$ limit, i.e., when the Drude plasma reduces to the electromagnetic vacuum, one finds that $\text{Re}\{\sigma_{ph,\omega}(\Omega)\} \to 0$. As expected, in such a case the bulk region cannot contribute to the photonic conductivity. As already discussed in Sect. II.A, for an unfilled cavity the dissipation is uniquely controlled by the interactions with the cavity walls (such an effect is not described by Eq. (30), which was derived considering periodic boundary conditions). The imaginary part of the conductivity cannot be evaluated in closed analytical form. It sets the relative phase of the photonic current relative to the mechanical acceleration.

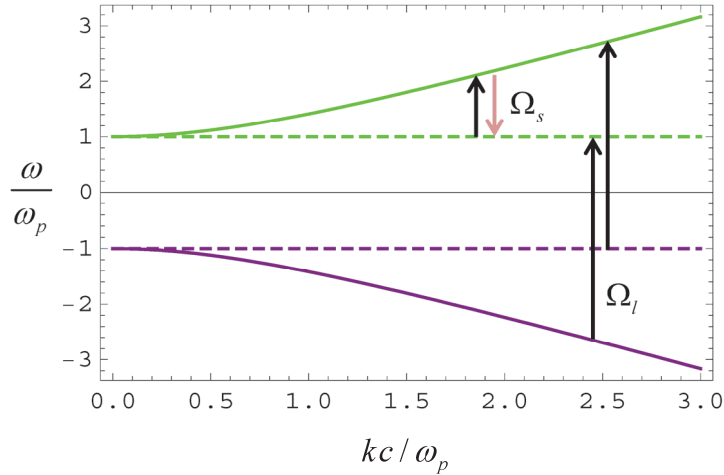

**Fig. 2** Band structure of the bulk material showing both the positive (green lines) and negative (purple lines) frequency bands and the interband transitions (vertical arrows) that determine the real part of the photonic conductivity. The dispersion of the transverse (longitudinal) plasmons is depicted with solid (dashed) lines, respectively. The black arrows (upward transitions) are associated with the generation of light quanta, which are eventually absorbed in the material (extraction of energy from the mechanical driving). The light red arrow (downward transition) is associated with the absorption of quanta from the thermal-light field, which is returned to the mechanical driver. The interactions between the positive frequency bands are possible for arbitrarily small

oscillation frequencies ($\Omega_s > 0$), whereas the interactions with positive and negative frequency bands are only feasible for $\Omega_l > 2\omega_p$.

The dissipation due to $\text{Re}\{\sigma_{ph,\omega}(\Omega)\} \neq 0$ can be traced back to interband transitions induced by the mechanical driving (see Appendix D), analogous to the loss induced by interband transitions in electronic systems. Photonic interband transitions have been previously discussed in different contexts [60-61]. All the four terms in Eq. (30) describe the interactions between volume plasmons and the transverse electromagnetic modes. The terms in the 1$^{st}$ line describe interactions between photonic bands with the same frequency sign (e.g., positive frequency oscillators), whereas the terms in the 2$^{nd}$ line describe interactions between photonic bands with opposite frequency signs (e.g., positive and negative frequency oscillators) (see Fig. 2). The latter are only feasible for sufficiently large $\Omega$, specifically for $\Omega > 2\omega_p$. All the upward transitions give rise to generation of light quanta (extraction of energy from the mechanical driving), whereas the downward transition gives rise to the absorption of light quanta (energy is returned to the mechanical driving).

The two terms of the first line in Eq. (30) describe the downward transition from a positive frequency transverse mode to a positive frequency longitudinal mode, and the opposite upward transition from a positive frequency longitudinal mode to a positive frequency transverse mode. The second process dominates so that the mechanical driving pumps energy into the transverse modes.

On the other hand, the two terms in the second line of Eq. (30) describe the transfer of energy from the branches with negative frequency to the branches with positive frequency, which always lead to irreversible light emission [16-20, 62].

The total conductivity $\boldsymbol{\sigma}_{ph}(\Omega) = \int_{0^+}^{+\infty} d\omega\, \boldsymbol{\sigma}_{ph,\omega}(\Omega)$ is given by,

$$\text{Re}\{\sigma_{ph}(\Omega)\} = \frac{1}{12\pi c}(\Omega + 2\omega_p)\frac{\sqrt{(\Omega+\omega_p)^2 - \omega_p^2}}{\Omega + \omega_p}\left[-\omega_p \mathcal{E}_{\Omega+\omega_p} + (\Omega+\omega_p)\mathcal{E}_{\omega_p}\right] \quad (31)$$
$$+ \frac{1}{12\pi c}(\Omega - 2\omega_p)\frac{\sqrt{(\Omega-\omega_p)^2 - \omega_p^2}}{\Omega - \omega_p}\left[+\omega_p \mathcal{E}_{\Omega-\omega_p} + (\Omega-\omega_p)\mathcal{E}_{\omega_p}\right].$$

Again, the term in the second line should be taken into account only when $\Omega > 2\omega_p$. In practical limit $\Omega \ll \omega_p$ the conductivity reduces to:

$$\text{Re}\{\sigma_{ph}(\Omega)\} \approx \frac{(\hbar\omega_p)^2}{k_B T \sinh^2\left(\frac{\hbar\omega_p}{2k_B T}\right)} \frac{\Omega}{12\pi c}\sqrt{\frac{\Omega\omega_p}{2}}, \qquad (\Omega \ll \omega_p). \quad (32)$$

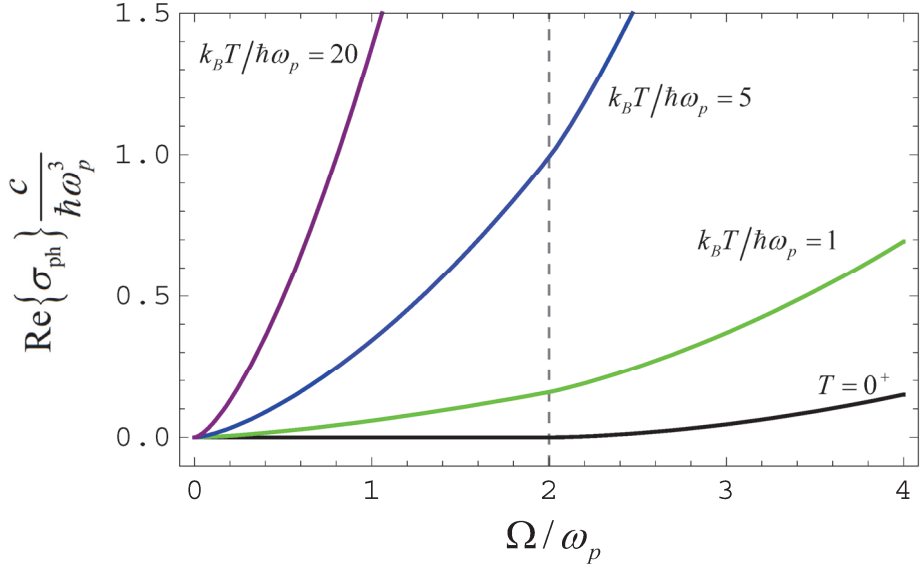

**Fig. 3** Photonic conductivity as a function of the normalized mechanical driving frequency $\Omega/\omega_p$ for different temperature values. In the $T = 0^+$ limit the real part of the photonic conductivity vanishes exactly for $\Omega < 2\omega_p$.

As illustrated in Fig. 3, the total conductivity $\sigma_{ph}(\Omega)$ is strictly positive, and hence in each oscillation cycle the mechanical driving transfers energy to the system. For $\Omega < 2\omega_p$ the extracted energy originates exclusively from the interaction between bulk plasmon and transverse mode oscillators with the same frequency sign (positive frequency oscillators).

Curiously, in the zero temperature limit, i.e., when the light fluctuations have a purely quantum origin, the contributions of the terms in the first line cancel out and thereby $\text{Re}\{\sigma_{ph}(\Omega)\} = 0$ for $\Omega < 2\omega_p$. In contrast, the dissipation arising from the interactions between positive frequency and negative frequency oscillators (for $\Omega > 2\omega_p$) is not suppressed for $T \to 0^+$. In fact, one obtains the following explicit formula for the zero temperature conductivity:

$$\text{Re}\{\boldsymbol{\sigma}_{ph}(\Omega)\}_{T=0^+} = \frac{\hbar \omega_p}{12\pi c}(\Omega - 2\omega_p)\sqrt{(\Omega - \omega_p)^2 - \omega_p^2}, \qquad \Omega > 2\omega_p. \qquad (33)$$

### C. Perturbation of the thermal-light energy in the cavity

Consider again the scenario of the previous subsection. The mechanical driving leads to an increase of the energy stored in the cavity. For $\Omega \ll \omega_p$ the generated light has a spectrum peaked at $\omega_p + \Omega$ [see Eq. (30)]. In the presence of some material dissipation, the delta-functions suffer some broadening resulting in a finite bandwidth $\Delta \omega$. For simplicity, we shall identify $\Delta \omega$ with the collision frequency $\Gamma$ of the material. The thermal energy in the undriven cavity in this spectral interval is $E_{th} = \mathcal{E}_{\omega_p + \Omega} n_{\omega_p + \Omega} V \Delta \omega$. Here, $V$ is the cavity volume, and $n_\omega$ is the density of photonic states (number of states per unit of frequency and per unit of volume). The density of states can be written as $n_\omega = \frac{1}{3\pi^2} \frac{d}{d\omega}\left(\omega\sqrt{\varepsilon\mu_0}\right)^3$, which for a Drude model with weak dissipation reduces to $n_\omega = \frac{1}{\pi^2} \frac{1}{c^3} \omega(\omega^2 - \omega_p^2)^{1/2}$ for $\omega > \omega_p$. Using this result, supposing always that $\Omega \ll \omega_p$, we find that the thermal energy is

$$E_{th} \approx \frac{1}{\pi^2} \frac{1}{c^3} \omega_p \sqrt{2\omega_p \Omega} \mathcal{E}_{\omega_p} V \Delta \omega.$$

The accelerated motion induces an excess of the thermal energy $\delta E_{th}$ in the relevant spectral range. The variation in time of $\delta E_{th}$ is ruled by $\langle p_{d \to EM} \rangle$ [Eq. (28)], as well as by the intrinsic dissipation of the material. Hence, it can be described by a differential equation of the type:

$$\frac{d}{dt}\delta E_{th} = \langle p_{d \to EM} \rangle - \Gamma \delta E_{th}. \qquad (34)$$

The second term on the right-hand side describes the dissipation in the material. We took into account that the damping rate near the plasma frequency is approximately the collision frequency $\Gamma$. Evidently, in a steady state we get $\delta E_{th} = \langle p_{d \to EM} \rangle / \Gamma$. In order to characterize the excess of thermal energy, next we evaluate $\delta E_{th} / E_{th}$. Using Eqs. (28), (32) and $\Delta \omega \approx \Gamma$, it can be shown that

$$\frac{\delta E_{th}}{E_{th}} \approx \frac{\pi}{12} \left( \frac{v_{max}}{c} \frac{\Omega}{\Gamma} \right)^2 \frac{\hbar \Omega}{k_B T} \left[ \sinh\left( \frac{\hbar \omega_p}{k_B T} \right) \right]^{-1}. \qquad (35)$$

In above $v_{max}$ is the peak velocity of the body in each oscillation cycle (the peak acceleration is $2a_0 = v_{max}\Omega$). It is interesting to consider the limit case $\hbar \omega_p \ll k_B T$. For example, at room temperature the inequality is satisfied up to terahertz frequencies, e.g., it is satisfied for typical semiconducting materials. In this case, the energy perturbation becomes independent of $\hbar, k_B, T$ and can be simply expressed as:

$$\frac{\delta E_{th}}{E_{th}} \approx \frac{\pi}{12} \left( \frac{v_{max}}{c} \right)^2 \left( \frac{\Omega}{\Gamma} \right)^2 \frac{\Omega}{\omega_p}. \qquad (36)$$

It is recalled that formula assumes that $\Omega \ll \omega_p$. To give an idea of the possibilities consider a semiconductor material with $\omega_p = 2\pi \times 2\,\text{THz}$ and collision frequency $\Gamma = 0.05\omega_p$ at room temperature ($T = 300K$). Consider the rather optimistic scenario where the peak velocity of the material is $v_{\max}/c = 10^{-3}$ and the driving frequency is $\Omega = 2\pi \times 20\,\text{GHz}$. Then, Eq. (36) predicts that $\delta E_{\text{th}}/E_{\text{th}} \sim 10^{-10}$ which is too small to be detected experimentally. Indeed, much like the dynamical Casimir effect, it appears that experimental verification is impractical when utilizing mechanical systems. The limited magnitude of the effect arises from the vast disparity in velocity and frequency scales between electromagnetic and mechanical parameters.

However, in principle, it is perfectly plausible to create an electronic synthetic motion that can mimic the required acceleration, e.g., with suitable spacetime modulations of the plasma. In such a case, it is possible to consider much larger $v_{\max}/c$ and $\Omega$. For example, using $v_{\max}/c = 0.1$ and $\Omega = 0.1\omega_p$ the relative energy variation becomes $\delta E_{\text{th}}/E_{\text{th}} \sim 10^{-3}$, which may be within experimental reach.

## V.  Photonic Hall Effect

When the frequency of the mechanical driving is vanishingly small, $\Omega \to 0$, i.e., for a constant acceleration, the spectral density (24) reduces to:

$$\boldsymbol{\sigma}_{\text{ph},\omega,ij}\bigg|_{\Omega=0} = \frac{-2\mathcal{E}_\omega}{V}\text{Re}\int d^3\mathbf{r}\int d^3\mathbf{r}'\,\text{Tr}\left\{\hat{\mathbf{S}}_i \cdot \partial_\omega \overline{\mathcal{G}}(\mathbf{r},\mathbf{r}',\omega)\cdot \hat{\mathbf{S}}_j \cdot \frac{1}{2\pi\omega}\left[\overline{\mathcal{G}}(\mathbf{r}',\mathbf{r},\omega) - \overline{\mathcal{G}}^\dagger(\mathbf{r}',\mathbf{r},\omega)\right]\right\} \quad (37)$$

In the following, we focus on the limit case where (in the spectral region of interest) the materials inside the box are nearly lossless. It will be shown that the constant acceleration of the box, may lead to a photonic Hall effect, such that the acceleration along a certain direction of space induces a photon current along a perpendicular direction. Furthermore, we will prove that for 2D-type platforms the photonic conductivity $\boldsymbol{\sigma}_{\text{ph},\omega}$ is quantized in the bulk band gaps and is determined by a topological invariant.

## A. Weak dissipation limit

Interestingly, in the weak dissipation limit (ideally for infinitesimal loss) $\boldsymbol{\sigma}_{ph,\omega}\big|_{\Omega=0}$ can be expressed in terms of the thermal-light angular momentum in the box in the absence of the mechanical driving. It is proven in Appendix E that:

$$\boldsymbol{\sigma}_{ph,\omega}\big|_{\substack{\Omega=0 \\ \text{weak dissipation}}} = c^2 \frac{1}{V} \mathcal{L}_\omega \times \mathbf{1}. \tag{38}$$

This result assumes that the box is terminated with opaque-type (non-cyclic) boundary conditions (e.g., perfectly conducting – PEC – walls). Here, $\mathcal{L}_\omega$ is the unilateral spectral density of the Abraham angular momentum ($\mathcal{L} = \int_{box} dV \frac{1}{c^2} \mathbf{r} \times \mathbf{S}(\mathbf{r})$), so that $\mathcal{L} = \int_0^\infty d\omega \mathcal{L}_\omega$. It is underlined that $\mathcal{L}_\omega$ is evaluated without the mechanical driving.

Evidently, $\mathcal{L}_\omega$ depends on the temperature and it can be expressed in terms of the system Green's function as in Eq. (E8). Moreover, $\mathcal{L}_\omega$ vanishes for reciprocal platforms (see Appendix E) [57, 63]. Thereby, we conclude that for weakly dissipative systems the photonic conductivity for a constant acceleration ($\Omega \to 0$) can be nontrivial only when the cavity contains nonreciprocal materials.

## B. Hall effect

Interestingly, Eq. (38) shows that $\boldsymbol{\sigma}_{ph,\omega}\big|_{\substack{\Omega=0 \\ \text{weak dissipation}}}$ is an anti-symmetric tensor such that the induced photon current is $\langle \mathbf{S}_{av,\omega} \rangle = -\frac{1}{V} \mathcal{L}_\omega \times \mathbf{a}$ for the constant acceleration $\mathbf{a}$. Thus, the component of the photon current with frequency $\omega$ is perpendicular to the acceleration $\mathbf{a}$ and to the fluctuation induced light angular momentum, analogous to the Hall effect in electronics. In physical terms, this implies that there is a net transverse light flow in the cavity. Curiously, from the perspective of the mechanical driving, there is no power

dissipation because $\mathbf{a} \cdot (\mathcal{L}_\omega \times \mathbf{a}) = 0$. We shall show in the following subsection –which is focused on topological systems– that the photonic Hall effect is a consequence of non-Hermitian asymmetric light-matter interactions. It is underlined that the Hall effect can only occur in nonreciprocal systems.

### C. Topological platforms and quantized photonic conductivity

It is relevant to consider 2D-type cavities, such that two dimensions of the box are much larger than the thickness along the $z$-direction: $L_x, L_y \gg L_z$. In such a case, for sufficiently long wavelengths, the light in the box can only flow along directions parallel to the *xoy* plane. Thus, the thermal-light angular momentum is directed along $z$: $\mathcal{L}_\omega = \mathcal{L}_\omega \hat{\mathbf{z}}$. For 2D-type systems, it is convenient to introduce $\langle \mathbf{g}_{av,\omega} \rangle_{2D} = \frac{1}{A} \frac{1}{c^2} \int_{box} \langle \mathbf{S}_\omega \rangle d^3 \mathbf{r}$ which is the spectral density of the light Abraham (kinetic) momentum in the box per unit of cross-sectional area ($A = L_x \times L_y$). From $\langle \mathbf{S}_{av,\omega} \rangle = -\frac{1}{V} \mathcal{L}_\omega \times \mathbf{a}$ it is evident that for weakly dissipative 2D-type systems $\langle \mathbf{g}_{av,\omega} \rangle_{2D}$ is linked to the constant-acceleration through a 2D photonic conductivity ($\boldsymbol{\sigma}_{2D,\omega}$) as follows (compare with Eq. (3)):

$$\langle \mathbf{g}_{av,\omega} \rangle_{2D} = -\frac{1}{c^4} \boldsymbol{\sigma}_{2D,\omega} \cdot \mathbf{a}, \qquad \text{with} \qquad \boldsymbol{\sigma}_{2D,\omega} = \frac{c^2 \mathcal{L}_\omega}{A} \hat{\mathbf{z}} \times \mathbf{1}. \qquad (39)$$

Thus, the spectral density of the angular momentum per unit of area determines the 2D photonic conductivity spectrum for $\Omega \to 0$. Importantly, it was shown in Refs. [64, 65] that in the bulk band-gaps of the (undriven) cavity $\mathcal{L}_\omega / A$ is quantized and is determined by the gap Chern number ($\mathcal{C}$) of the bulk material band-gap as follows:

$$\left. \frac{\mathcal{L}_\omega}{A} \right|_{band-gap} = -\frac{\mathcal{C}}{\pi c^2} \mathcal{E}_\omega. \qquad (40)$$

Note that $\mathcal{L}_\omega / A$ has unities of mass. Thus, in the band-gaps, the 2D-photonic conductivity is quantized in units of $\mathcal{E}_\omega / \pi$:

$$\boldsymbol{\sigma}_{2D,\omega} = -\frac{\mathcal{C}}{\pi} \mathcal{E}_\omega \hat{\mathbf{z}} \times \mathbf{1}, \qquad \text{(band-gaps)}. \qquad (41)$$

For sufficiently large temperatures, $\frac{\hbar\omega}{2k_B T} \ll 1$, the energy $\mathcal{E}_\omega$ is determined simply by $\mathcal{E}_\omega \approx k_B T$. Evidently, $\boldsymbol{\sigma}_{2D,\omega}$ is associated with a quantum-photonic Hall effect.

The origin of the quantum-photonic Hall effect may be understood as follows. From Sect. II.C (see also Appendix B), the constant acceleration of the box leads on nonuniform distribution of the thermal-light energy inside the box, such the energy density is larger near the wall opposite to the direction of $\mathbf{a}$ (wall $X = -L_x / 2$ for an acceleration along $+x$). In a band-gap of a topological material, the thermal-light energy is transported by unidirectional topological edge states that circulate around the cavity walls [64, 65]. As the Poynting vector of a mode is proportional to the energy density, the enunciated properties lead to the picture shown in Fig. 4, which represents the Poynting vector lines for a 2D-type cavity accelerated along the +x-direction. The sketch assumes that in the considered band-gap the cavity supports a single topological edge-state that circulates in the counter-clockwise direction ($\mathcal{C} = -1$, see [64, 65]). Consistent with the previous discussion, the Poynting vector has the largest intensity (largest arrows) near the wall $X = -L_x / 2$ and the smallest intensity near the opposite wall $X = +L_x / 2$ (smallest arrows). In the side walls, the Poynting vector varies exponentially. For realistic values of the acceleration, the Poynting vector amplitude change is rather tiny and due to this reason in practice the exponential variation can be approximated by a linear variation with a very small slope. Consistent with Appendix B, the decay rate is independent of direction of propagation of the edge mode. Hence, the Poynting vector (or the light momentum) averaged over the transverse cross-section is determined only by the contributions of the walls $X = \pm L_x / 2$. Evidently, in the conditions of Fig. 4, the averaged light momentum is directed along the $-y$-direction, in agreement with Eqs. (39) and (41). Note that the Poynting vector lines in Fig. 4 include both the thermal contribution (without the driving) and the perturbation due to the mechanical driving. The Poynting vector lines without the mechanical driving are nontrivial [57], but their volume average vanishes exactly [65] (i.e., the expectation of the light momentum in the cavity without the driving vanishes exactly). Note that $\langle \mathbf{g}_{av,\omega} \rangle_{2D}$ given by Eq. (39) depends only on the perturbation of the Poynting vector lines.

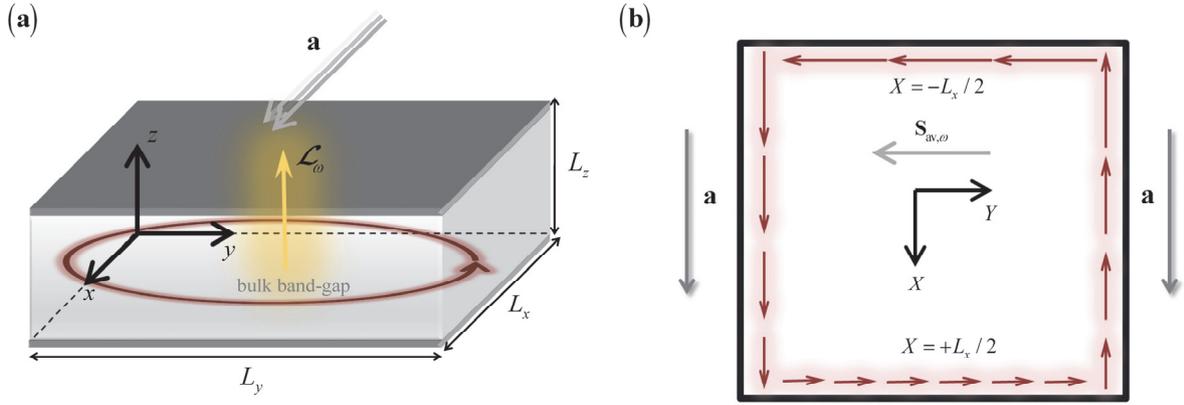

**Fig. 4** (a) Illustration of a 2D-type topological cavity subject to a constant acceleration $\mathbf{a}$. The bulk region does not support electromagnetic states at the frequency range of interest. For topological systems, the thermal light spectrum inside the cavity is determined by the number of topological states. Here, it is assumed that the topological gap is associated with a single counter-clockwise unidirectional edge state ($\mathcal{C} = -1$). (b) Sketch of the expectation of the Poynting vector lines when the cavity is accelerated along the $+x$-direction. Due to "electromagnetic inertia", the thermal energy associated with the perturbed unidirectional edge state tends to be concentrated on the wall $X = -L_x / 2$. This originates a photonic Hall effect, with the averaged Poynting vector expectation oriented along $-y$.

The non-uniformity of the Poynting vector lines along the sidewalls ($Y = \pm L_y / 2$) implies that the fields near these walls must be pumped by the accelerated motion. The edge wave near the wall $Y = +L_y / 2$ ($Y = -L_y / 2$) draws energy from (returns energy to) the mechanical driver, respectively. On the overall, this process does not require any transfer of energy from the driving ($\langle p_{\text{d} \to \text{EM}} \rangle = 0$) because the contributions of the two opposite walls cancel out. The phenomenon is reminiscent of the photon thermal Hall effect [68], where a longitudinal temperature gradient (mimicking the acceleration along *x*) induces a transverse heat flux between two reservoirs initially at the same temperature.

In principle, the described effect may be most easily experimentally verified using time-crystals that mimic the accelerated motion of the medium and of the cavity walls. In such a context, it is expected that the thermally generated heat current (in the gap spectral region) is stronger near the wall at $X = -L_x/2$ and weaker near the wall $X = +L_x/2$. Similar to the mechanical problem, the synthetic (electronic) driving of the medium is expected to disrupt the balance between the radiation and dissipation of the noise sources in the medium, such that on average the circuitry near the wall $Y = +L_y/2$ pumps energy into the cavity leading to the amplification of the edge mode, whereas the circuitry near the opposite wall $Y = -L_y/2$, extracts energy from the edge mode leading to its exponential decay. On average, the driving circuit does not inject or remove energy into the system.

## VI.     Conclusion

In summary, it was shown that one may formally define a "photonic conductivity" that describes the thermal-light momentum linear response to the mechanical oscillations of a medium. Similar to the electronic case, the real (Hermitian) part of the photonic conductivity determines a dissipative response, which is associated with an irreversible transfer of energy from the mechanical driving to the electromagnetic field. The energy extracted from the mechanical driving is ultimately dissipated in the form of heat. We determined the photonic conductivity for a weakly-dissipative dispersive plasma with the initial field distribution determined by the Bose-Einstein statistics. Our results show that the irreversible light emission is originated by the upward interband transitions triggered by the mechanical driving.

Our findings reveal that the photonic conductivity of a weakly dissipative material enclosed in a cavity, under constant acceleration, is determined by the thermal-light angular momentum spectrum. The mechanical driving originates a photonic Hall effect where the induced light momentum is perpendicular to the constant acceleration. The photonic Hall effect arises due to strongly asymmetric light-matter interactions. Specifically, the edge mode near one of the cavity walls draws energy from the mechanical system, whereas the edge mode on the opposite wall returns energy to the mechanical system, similar to a transverse temperature gradient.

Notably, for 2D-type cavities, the photonic conductivity under constant acceleration is quantized in the gaps of the bulk region, with the quantum of photonic conductivity being determined by the Chern number of the 2D-photonic system. This quantization establishes a precise parallel with the electronic quantum Hall effect. Moreover, these results can be extended to certain classes of spacetime-modulated systems, where the mechanical driving is replaced by the electronic time modulation of the material parameters.

**Acknowledgements:** This work is supported in part by the Institution of Engineering and Technology (IET), by the Simons Foundation, and by Fundação para a Ciência e a Tecnologia and Instituto de Telecomunicações under project UIDB/50008/2020.

## Appendix A: Coordinate transformations

We start from the Maxwell's equations in the laboratory reference frame:

$$\nabla \times \mathbf{E} = -\partial_t \mathbf{B}, \qquad \nabla \times \frac{\mathbf{B}}{\mu_0} = \varepsilon_0 \partial_t \mathbf{E} + \mathbf{j}. \tag{A1}$$

Here, $\mathbf{j}$ includes all the polarization currents of the accelerated medium. The corresponding electric charge density is $\rho$. The trajectory of the medium (regarded as a rigid body) is determined by $\mathbf{r}_0(t)$. The instantaneous velocity of the medium with respect to the laboratory frame is $\mathbf{v} = d\mathbf{r}_0/dt$ and the acceleration is $\mathbf{a} = d\mathbf{v}/dt$. For simplicity, it is supposed that that at initial time $\mathbf{r}_0(0) = 0$ and so $\mathbf{r}_0(t)$ is the rigid body displacement due to the motion.

We consider a field transformation of the type:

$$\mathbf{E}_{co} = \mathbf{E} + \mathbf{v} \times \mathbf{B}, \qquad \mathbf{B}_{co} = \mathbf{B} - \frac{1}{c^2} \mathbf{v} \times \mathbf{E}, \qquad (A2a)$$

$$\mathbf{j}_{co} = \mathbf{j} - \rho \mathbf{v}, \qquad \rho_{co} = \rho - \mathbf{j} \cdot \mathbf{v}/c^2. \qquad (A2b)$$

Neglecting terms that are on the order of $v^2$ ($o(v^2)$ approximation), the inverse transformation is $\mathbf{E} = \mathbf{E}_{co} - \mathbf{v} \times \mathbf{B}_{co}$, $\mathbf{B} = \mathbf{B}_{co} + \frac{1}{c^2} \mathbf{v} \times \mathbf{E}_{co}$, and $\mathbf{j} = \mathbf{j}_{co} + \rho_{co} \mathbf{v}$, $\rho = \rho_{co} + \mathbf{j}_{co} \cdot \mathbf{v}/c^2$. Within a $o(v^2)$ approximation, the fields with subscript "co" may be regarded as the fields in the frame that moves with velocity $\mathbf{v}$ with respect to the laboratory frame [55]. The transformation leads to:

$$\nabla \times (\mathbf{E}_{co} - \mathbf{v} \times \mathbf{B}_{co}) + \left(\frac{1}{c^2}\mathbf{a} + \frac{1}{c^2}\mathbf{v}\partial_t\right) \times \mathbf{E}_{co} = -\partial_t \mathbf{B}_{co}, \qquad (A3)$$

$$\nabla \times \left(\frac{\mathbf{B}_{co}}{\mu_0} + \mathbf{v} \times \varepsilon_0 \mathbf{E}_{co}\right) + \left(\frac{1}{c^2}\mathbf{a} + \frac{1}{c^2}\mathbf{v}\partial_t\right) \times \frac{\mathbf{B}_{co}}{\mu_0} = \varepsilon_0 \partial_t \mathbf{E}_{co} + \mathbf{j}_{co} + \rho_{co} \mathbf{v}. \qquad (A4)$$

Next, we consider a Galilean-type coordinate transformation:

$$\mathbf{r} = \mathbf{r}_0(\tau) + \mathbf{R}, \qquad t = \tau. \qquad (A5)$$

The coordinate transformation implies that:

$$\nabla_\mathbf{R} = \nabla, \qquad \partial_\tau = \mathbf{v} \cdot \nabla + \partial_t. \qquad (A6)$$

Neglecting terms that are $o(v^2)$ we get:

$$\nabla_{\mathbf{R}} \times \left(\mathbf{E}_{co} - \mathbf{v} \times \mathbf{B}_{co}\right) + \left(\frac{1}{c^2}\mathbf{a} + \frac{1}{c^2}\mathbf{v}\partial_\tau\right) \times \mathbf{E}_{co} = -\left(\partial_\tau - \mathbf{v} \cdot \nabla_{\mathbf{R}}\right)\mathbf{B}_{co}, \tag{A7a}$$

$$\nabla_{\mathbf{R}} \times \left(\frac{\mathbf{B}_{co}}{\mu_0} + \mathbf{v} \times \varepsilon_0 \mathbf{E}_{co}\right) + \left(\frac{1}{c^2}\mathbf{a} + \frac{1}{c^2}\mathbf{v}\partial_\tau\right) \times \frac{\mathbf{B}_{co}}{\mu_0} = \left(\partial_\tau - \mathbf{v} \cdot \nabla_{\mathbf{R}}\right)\varepsilon_0 \mathbf{E}_{co} + \mathbf{j}_{co} + \rho_{co}\mathbf{v}. \tag{A7b}$$

Taking into account that $\nabla_{\mathbf{R}} \cdot \mathbf{B}_{co} = o(v)$, and $\nabla_{\mathbf{R}} \cdot \mathbf{E}_{co} = \rho_{co}/\varepsilon_0 + o(v)$, we see that

$$\nabla_{\mathbf{R}} \times \left(\mathbf{v} \times \mathbf{B}_{co}\right) = \mathbf{v}\nabla_{\mathbf{R}} \cdot \mathbf{B}_{co} - \left(\mathbf{v} \cdot \nabla_{\mathbf{R}}\right)\mathbf{B}_{co} = -\left(\mathbf{v} \cdot \nabla_{\mathbf{R}}\right)\mathbf{B}_{co} + o(v^2) \quad \text{and that}$$

$\nabla_{\mathbf{R}} \times \left(\mathbf{v} \times \mathbf{E}_{co}\right) = \mathbf{v}\rho_{co}/\varepsilon_0 - \left(\mathbf{v} \cdot \nabla_{\mathbf{R}}\right)\mathbf{E}_{co} + o(v^2)$. This finally shows that:

$$\nabla_{\mathbf{R}} \times \mathbf{E}_{co} + \left(\frac{1}{c^2}\mathbf{a} + \frac{1}{c^2}\mathbf{v}\partial_\tau\right) \times \mathbf{E}_{co} = -\partial_\tau \mathbf{B}_{co}, \tag{A8a}$$

$$\nabla_{\mathbf{R}} \times \frac{\mathbf{B}_{co}}{\mu_0} + \left(\frac{1}{c^2}\mathbf{a} + \frac{1}{c^2}\mathbf{v}\partial_\tau\right) \times \frac{\mathbf{B}_{co}}{\mu_0} = \varepsilon_0 \partial_\tau \mathbf{E}_{co} + \mathbf{j}_{co}. \tag{A8b}$$

It is useful to note that $\tau = t$ may be regarded as the proper time measured by an observer (e.g., a clock attached to the point with coordinates $\mathbf{r} = \mathbf{R}$ at initial time when $\mathbf{r}_0(0) = 0$) that follows the same trajectory as the moving body. In fact, as a proper time (infinitesimal) interval is determined by $d\tau = \sqrt{(dt)^2 - \frac{1}{c^2}(\Delta \mathbf{r})^2} = \sqrt{(dt)^2 - \frac{1}{c^2}(\Delta \mathbf{r}_0(t))^2}$, it follows that the proper time is $\tau = \int_0^t dt\sqrt{1 - \frac{\mathbf{v}_0(t) \cdot \mathbf{v}_0(t)}{c^2}} \approx t + o(v^2)$. For a local material, the dynamics of the polarization currents $\mathbf{j}_{pol,co} = \frac{\partial}{\partial \tau}\mathbf{P}_{co}$ is controlled by a differential equation that depends on the *proper time* ($\mathbf{P}_{co}$ is the polarization vector in the co-moving frame). For example, for a Drude plasma (e.g., a metal) the current should satisfy $\partial_\tau \mathbf{j}_{pol,co} + \Gamma \mathbf{j}_{pol,co} = \varepsilon_0 \omega_p^2 \mathbf{E}_{co}$, with $\Gamma, \omega_p$ the collision and plasma frequencies, respectively. This observation shows that the polarization currents $\mathbf{j}_{pol,co} = \frac{\partial}{\partial \tau}\mathbf{P}_{co}$ are controlled by exactly the same differential equations in time as in the static case (without any motion). Hence, Eq. (A8) can be written in the compact form as:

$$\hat{L}(-i\nabla_\mathbf{R}) \cdot \mathbf{f}_{co} + \underbrace{\left(-i\frac{\mathbf{a}}{c^2} \cdot \hat{\mathbf{S}} - i\frac{\mathbf{v}}{c^2} \cdot \hat{\mathbf{S}} \partial_\tau \right)}_{\hat{L}_{int}} \cdot \mathbf{f}_{co} = i\frac{\partial}{\partial \tau} \mathbf{g}_{co} + i\mathbf{j}_{ext,co} \ . \tag{A9}$$

Here, $\mathbf{f}_{co} = (\mathbf{E}_{co} \ \ \mathbf{H}_{co})^T$, $\mathbf{g}_{co} = (\mathbf{D}_{co} \ \ \mathbf{B}_{co})^T$, where by definition $\mathbf{H}_{co} = \mathbf{B}_{co}/\mu_0$ and $\mathbf{D}_{co} = \varepsilon_0 \mathbf{E}_{co} + \mathbf{P}_{co}$. For a dispersive dielectric, the field $\mathbf{D}_{co}(\mathbf{R},\tau)$ is linked to $\mathbf{E}_{co}(\mathbf{R},\tau)$ exactly by the same differential equations in time, as in the rest case (or alternatively, by a time convolution with a suitable kernel). Moreover, $\mathbf{j}_{ext,co}$ represents the external excitation currents (i.e., the currents that are not part of the medium). The differential operator $\hat{L}$ is given by

$$\hat{L}(-i\nabla_\mathbf{R}) = \begin{pmatrix} \mathbf{0}_{3\times 3} & i\nabla_\mathbf{R} \times \mathbf{1}_{3\times 3} \\ -i\nabla_\mathbf{R} \times \mathbf{1}_{3\times 3} & \mathbf{0}_{3\times 3} \end{pmatrix} . \tag{A10}$$

On the other hand, the interaction term $\hat{L}_{int}$ due to the medium acceleration is written in terms of the matrices

$$\hat{\mathbf{S}}_i = \begin{pmatrix} \mathbf{0}_{3\times 3} & -\hat{\mathbf{u}}_i \times \mathbf{1}_{3\times 3} \\ \hat{\mathbf{u}}_i \times \mathbf{1}_{3\times 3} & \mathbf{0}_{3\times 3} \end{pmatrix}, \qquad i = 1,2,3, \tag{A11}$$

with $\hat{\mathbf{u}}_i$ a unit vector along the $i$-th space direction. By definition, we have $\mathbf{a} \cdot \hat{\mathbf{S}} = a_1 \hat{\mathbf{S}}_1 + a_2 \hat{\mathbf{S}}_2 + a_3 \hat{\mathbf{S}}_3 = \begin{pmatrix} \mathbf{0}_{3\times 3} & -\mathbf{a} \times \mathbf{1}_{3\times 3} \\ \mathbf{a} \times \mathbf{1}_{3\times 3} & \mathbf{0}_{3\times 3} \end{pmatrix}$. Note that in absence of the mechanical driving ($\mathbf{v} = 0 = \mathbf{a}$) Eq. (A9) reduces the Maxwell's equations in the relevant dispersive material structure.

## Appendix B: Electromagnetic modes in the limit $\Omega \to 0$

Here, we study how the acceleration affects the electromagnetic modes of the cavity in the limit $\Omega \to 0$ (constant acceleration, $\mathbf{a} = a\hat{\mathbf{x}}$). It is supposed that in the time range of interest the velocity of the system is negligible. Then, the source-free master equation (A9) [with $\mathbf{j}_{\text{ext,co}} = 0$] in the $\mathbf{R}, \tau$ coordinates is equivalent to:

$$\nabla_{\mathbf{R}} \times \mathbf{E}_{\text{co}} + \frac{1}{c^2}\mathbf{a} \times \mathbf{E}_{\text{co}} = -\partial_\tau \mathbf{B}_{\text{co}}, \tag{B1a}$$

$$\nabla_{\mathbf{R}} \times \mathbf{H}_{\text{co}} + \frac{1}{c^2}\mathbf{a} \times \mathbf{H}_{\text{co}} = \partial_\tau \mathbf{D}_{\text{co}}. \tag{B1b}$$

It is underlined that without the mechanical driving the above equations reduce to the Maxwell's equations in the undriven system (with all the material structures at rest). It is straightforward to construct the modes of the above equation (with $\mathbf{a} = const.$) from the modes of the undriven system. In fact, it is clear that the solutions with $\mathbf{a} \neq 0$ are related to the solutions with $\mathbf{a} = 0$ as:

$$\mathbf{E}_{\text{co}} = \mathbf{E}_{\mathbf{a}=0} e^{-\frac{\mathbf{a}}{c^2}\cdot\mathbf{R}}. \qquad \mathbf{H}_{\text{co}} = \mathbf{H}_{\mathbf{a}=0} e^{-\frac{\mathbf{a}}{c^2}\cdot\mathbf{R}}. \tag{B2}$$

Thus, in the co-moving frame coordinates electromagnetic modes of the accelerated system are exactly the same as the modes of the undriven system, apart from the exponential factor $e^{-\frac{a}{c^2}X}$. In other words, the accelerated motion leads to an exponential decay along $+x$ and to the concentration of energy on the back wall of the box. Note that the exponential decay is insensitive to the direction of propagation of the wave. Clearly, the exponential factor implies that both the fields and the Poynting vector in the laboratory frame coordinates decay exponentially along $+x$.

# Appendix C: Modal expansion of the photonic conductivity in the lossless limit

In the following, we obtain a modal expansion of the photonic conductivity [Eq. (23)] in the limit of weak dissipation. In this limit, the Green's function can be expanded into the electromagnetic modes of the system, $\mathbf{f}_n(\mathbf{r}) = (\mathbf{E}_n \quad \mathbf{H}_n)^T$, with real-valued eigenfrequencies $\omega_n$ [66, 67]:

$$\overline{\mathcal{G}}(\mathbf{r}|\mathbf{r}',\omega) = \frac{\omega}{2}\sum_n \frac{1}{\omega_n - \omega}\mathbf{f}_n(\mathbf{r})\otimes\mathbf{f}_n^*(\mathbf{r}'). \tag{C1}$$

Note that the Green's function definition here differs slightly from the definition of Refs. [66, 67]. The eigenmodes are normalized as [66, 67]:

$$\frac{1}{2}\int_V d^3\mathbf{r}\, \mathbf{f}_n^*(\mathbf{r})\cdot \partial_\omega\left[\omega\mathbf{M}(\omega,\mathbf{r})\right]_{\omega=\omega_n}\cdot \mathbf{f}_n(\mathbf{r}) = 1, \tag{C2}$$

where $\mathbf{M}(\omega,\mathbf{r})$ is the material matrix [Eq. (11)].

It is straightforward to show that [57, 66]:

$$\frac{1}{2\pi}\left[\frac{\overline{\mathcal{G}}(\mathbf{r}',\mathbf{r},\omega)}{\omega} - \left[\frac{\overline{\mathcal{G}}(\mathbf{r},\mathbf{r}',\omega)}{\omega}\right]^\dagger\right]_{\omega+0^+ i} = \frac{i}{2}\sum_n \delta(\omega-\omega_n)\mathbf{f}_n(\mathbf{r}')\otimes\mathbf{f}_n^*(\mathbf{r}). \tag{C3}$$

Using this result in the conductivity spectral density [Eq. (24), evaluated with $\omega \to \omega + 0^+ i$, i.e., a small positive imaginary part] one finds that:

$$\boldsymbol{\sigma}_{\text{ph},\omega,ij}(\Omega) = \frac{-\mathcal{E}_\omega}{V}\int d^3\mathbf{r}\int d^3\mathbf{r}'$$

$$\text{Tr}\left\{\hat{\mathbf{S}}_i\cdot\left[\frac{1}{2}\sum_m \frac{1}{\omega-\omega_m}\frac{\omega_m}{\omega+\Omega-\omega_m}\mathbf{f}_m(\mathbf{r})\otimes\mathbf{f}_m^*(\mathbf{r}')\right]_{\omega+0^+ i}\cdot\hat{\mathbf{S}}_j\cdot\frac{i}{2}\sum_m \delta(\omega-\omega_n)\mathbf{f}_n(\mathbf{r}')\otimes\mathbf{f}_n^*(\mathbf{r})\right\}$$

$$+\text{Tr}\left\{\hat{\mathbf{S}}_i\cdot\left[\frac{1}{2}\sum_m \frac{1}{\omega-\omega_m}\frac{\omega_m}{\omega-\Omega-\omega_m}\mathbf{f}_m(\mathbf{r})\otimes\mathbf{f}_m^*(\mathbf{r}')\right]_{\omega+0^+ i}\cdot\hat{\mathbf{S}}_j\cdot\frac{i}{2}\sum_m \delta(\omega-\omega_n)\mathbf{f}_n(\mathbf{r}')\otimes\mathbf{f}_n^*(\mathbf{r})\right\}^*$$

$$\tag{C4}$$

This can also be written as:

$$\boldsymbol{\sigma}_{\mathrm{ph},\omega,ij}(\Omega) = \frac{-\mathcal{E}_\omega}{V} i \sum_{m,n} \delta(\omega - \omega_n) \left[ \frac{1}{\omega_{nm} + 0^+ i} \frac{\omega_m}{\omega_{nm} + (\Omega + 0^+ i)} \langle n|\hat{\mathbf{S}}_i|m\rangle\langle m|\hat{\mathbf{S}}_j|n\rangle + \right. \\ \left. - \frac{1}{\omega_{nm} - 0^+ i} \frac{\omega_m}{\omega_{nm} - (\Omega + 0^+ i)} \langle m|\hat{\mathbf{S}}_i|n\rangle\langle n|\hat{\mathbf{S}}_j|m\rangle \right] \quad (C5)$$

with $\omega_{nm} = \omega_n - \omega_m$ and $\langle m|\hat{\mathbf{S}}_i|n\rangle = \frac{1}{2}\int \mathbf{f}_m^* \cdot \hat{\mathbf{S}}_i \cdot \mathbf{f}_n \, d^3\mathbf{r}$. Noting that the $\delta$-functions with negative frequency can be dropped due to the unilateral nature of the spectral density and that the terms with $m = n$ cancel out, one finally finds:

$$\boldsymbol{\sigma}_{\mathrm{ph},\omega,ij}(\Omega) = i\frac{\mathcal{E}_\omega}{V} \sum_{\substack{\omega_m \neq \omega_n, \\ \omega_n > 0}} \delta(\omega - \omega_n) \frac{\omega_m}{\omega_{mn}} \left[ \frac{1}{(\Omega + 0^+ i) - \omega_{mn}} \langle n|\hat{\mathbf{S}}_i|m\rangle\langle m|\hat{\mathbf{S}}_j|n\rangle + \right. \\ \left. + \frac{1}{(\Omega + 0^+ i) + \omega_{mn}} \langle m|\hat{\mathbf{S}}_i|n\rangle\langle n|\hat{\mathbf{S}}_j|m\rangle \right] \quad (C6)$$

with $\omega_{mn} = \omega_m - \omega_n$.

Let us now consider the limit $\Omega \to 0$, i.e., a constant acceleration. In that case, the spectral density of the conductivity reduces to:

$$\boldsymbol{\sigma}_{\mathrm{ph},\omega,ij}\big|_{\Omega=0} = i\frac{\mathcal{E}_\omega}{V} \sum_{\substack{\omega_m \neq \omega_n, \\ \omega_n > 0}} \delta(\omega - \omega_n) \frac{\omega_m}{\omega_{mn}^2} \left[ -\langle n|\hat{\mathbf{S}}_i|m\rangle\langle m|\hat{\mathbf{S}}_j|n\rangle + \langle m|\hat{\mathbf{S}}_i|n\rangle\langle n|\hat{\mathbf{S}}_j|m\rangle \right] \quad (C7)$$

The above tensor is clearly anti-symmetric and real-valued.

## Appendix D: Photonic conductivity of a Drude plasma

Here, we derive the photonic conductivity of a uniform Drude plasma. For simplicity, we neglect the effect of the boundary walls, i.e., we only take into account the bulk interactions. This can be done by considering periodic boundary conditions in a large volume with dimensions $L_x \times L_y \times L_z$. Evidently, within such approximation the system becomes effectively invariant to continuous translations of space, and thereby the Green's function is of the form $\overline{\mathcal{G}}(\mathbf{r},\mathbf{r}',\omega) = \overline{\mathcal{G}}(\mathbf{r}-\mathbf{r}',\omega)$.

To proceed, it is convenient to switch to the spectral domain and use

$$\overline{\mathcal{G}}(\mathbf{r},\mathbf{r}',\omega) = \frac{1}{(2\pi)^3}\int d^3\mathbf{k}\, \overline{\mathcal{G}}_\mathbf{k}(\omega) e^{i\mathbf{k}\cdot(\mathbf{r}-\mathbf{r}')}, \tag{D1}$$

where $\overline{\mathcal{G}}_\mathbf{k}(\omega) = \omega\left[\mathbf{k}\cdot\hat{\mathbf{S}} - \omega\mathbf{M}(\omega)\right]^{-1}$ is the Fourier transform of $\overline{\mathcal{G}}(\mathbf{r},\omega)$ (see Eq. (10)). Substituting the above formula into Eq. (24b) it can be shown after straightforward manipulations that the bilateral photonic conductivity spectral density can be expressed as:

$$\boldsymbol{\sigma}^b_{\text{ph},\omega,ij}(\Omega) = -\frac{\mathcal{E}_\omega}{(2\pi)^3}\int d^3\mathbf{k}\, \text{Tr}\left\{\hat{\mathbf{S}}_i\cdot\frac{1}{\Omega}\left[\overline{\mathcal{G}}_\mathbf{k}(\omega+\Omega) - \overline{\mathcal{G}}_\mathbf{k}(\omega)\right]\cdot\hat{\mathbf{S}}_j\cdot\frac{i}{\pi}\text{Im}\left\{\frac{\overline{\mathcal{G}}_\mathbf{k}(\omega)}{\omega}\right\}\right\} \tag{D2}$$

We took into account that for an isotropic dispersive dielectric $\overline{\mathcal{G}}_\mathbf{k}(\omega)$ has transpose symmetry.

In order to derive an explicit analytical formula for $\boldsymbol{\sigma}^b_{\text{ph},\omega,ij}(\Omega)$, next we consider the weak dissipation limit. In such a case the function $\frac{\overline{\mathcal{G}}_\mathbf{k}(\omega)}{\omega}$ can be expanded into partial fractions. The poles of $\frac{\overline{\mathcal{G}}_\mathbf{k}(\omega)}{\omega}$ are determined by the dispersion of the electromagnetic modes of the bulk material, and in addition by the pole at $\omega = 0$. A lossless Drude plasma supports transverse electromagnetic modes with dispersion $\omega_k = \pm\sqrt{\omega_p^2 + c^2 k^2}$, and longitudinal modes (bulk plasmons) with dispersion $\pm\omega_p$. Note that one needs to consider both positive frequency and negative frequency modes in the partial-fraction expansion. Thus, $\frac{\overline{\mathcal{G}}_\mathbf{k}(\omega)}{\omega}$ has a partial fraction expansion of the form:

$$\frac{\overline{\mathcal{G}}_{\mathbf{k}}(\omega)}{\omega} = \frac{\mathbf{R}_0}{\omega} + \mathbf{R}_{L\mathbf{k}}\left[\frac{1}{\omega-\omega_p} + \frac{1}{\omega+\omega_p}\right] + \left[\frac{\mathbf{R}_{T\mathbf{k}}^+}{\omega-\omega_k} + \frac{\mathbf{R}_{T\mathbf{k}}^-}{\omega+\omega_k}\right]. \tag{D3}$$

In the above, $\mathbf{R}_0, \mathbf{R}_{L\mathbf{k}}, \mathbf{R}_{T\mathbf{k}}^\pm$ are the residues of the Green's function, for example

$\mathbf{R}_{L\mathbf{k}} = \lim_{\omega \to \omega_p}(\omega-\omega_p)\frac{\overline{\mathcal{G}}_{\mathbf{k}}(\omega)}{\omega}$, etc. Explicit calculations show that:

$$\mathbf{R}_{T\mathbf{k}}^s = \begin{pmatrix} \frac{1}{2\varepsilon_0 k^2}\left(\mathbf{k}\otimes\mathbf{k} - \mathbf{1}_{3\times 3}k^2\right) & s\frac{c^2}{2\omega_k}\mathbf{k}\times\mathbf{1}_{3\times 3} \\ -s\frac{c^2}{2\omega_k}\mathbf{k}\times\mathbf{1}_{3\times 3} & \frac{\varepsilon_0 c^4}{2\omega_k^2}\left(\mathbf{k}\otimes\mathbf{k} - \mathbf{1}_{3\times 3}k^2\right) \end{pmatrix}, \quad s = \pm \tag{D4a}$$

$$\mathbf{R}_{L\mathbf{k}} = \begin{pmatrix} \frac{-1}{2\varepsilon_0 k^2}\mathbf{k}\otimes\mathbf{k} & \mathbf{0}_{3\times 3} \\ \mathbf{0}_{3\times 3} & \mathbf{0}_{3\times 3} \end{pmatrix}, \tag{D4b}$$

$$\mathbf{R}_0 = \begin{pmatrix} \mathbf{0}_{3\times 3} & \mathbf{0}_{3\times 3} \\ \mathbf{0}_{3\times 3} & -\frac{\varepsilon_0 c^4}{\omega_k^2}\left(\mathbf{k}\otimes\mathbf{k} + \mathbf{1}_{3\times 3}\frac{\omega_p^2}{c^2}\right) \end{pmatrix}, \tag{D4c}$$

Using the partial fraction expansion, it is possible to show that:

$$\frac{1}{\Omega}\left[\overline{\mathcal{G}}_{\mathbf{k}}(\omega+\Omega) - \overline{\mathcal{G}}_{\mathbf{k}}(\omega)\right] = \mathbf{R}_{L\mathbf{k}} A_L(\omega) + \mathbf{R}_{T\mathbf{k}}^+ A_{T\mathbf{k}}^+(\omega) + \mathbf{R}_{T\mathbf{k}}^+ A_{T\mathbf{k}}^-(\omega) \tag{D5a}$$

$$A_L(\omega) = \frac{-\omega_p}{(\omega-\omega_p)(\omega+\Omega-\omega_p)} + \frac{\omega_p}{(\omega+\omega_p)(\omega+\Omega+\omega_p)}, \tag{D5b}$$

$$A_{T\mathbf{k}}^\pm(\omega) = \frac{-(\pm\omega_k)}{(\omega-(\pm\omega_k))(\omega+\Omega-(\pm\omega_k))}. \tag{D5c}$$

On the other hand, considering that $\omega \to \omega + 0^+ i$ is in the upper-half frequency plane, one finds that:

$$\frac{1}{\pi}\mathrm{Im}\left\{\frac{\overline{\mathcal{G}}_{\mathbf{k}}(\omega)}{\omega}\right\} = -\mathbf{R}_{L\mathbf{k}}\left[\delta(\omega-\omega_p) + \delta(\omega+\omega_p)\right] - \left[\mathbf{R}_{T\mathbf{k}}^+ \delta(\omega-\omega_k) + \mathbf{R}_{T\mathbf{k}}^- \delta(\omega+\omega_k)\right]. \tag{D6}$$

In the above formula, we dropped the contribution from $\mathbf{R}_0$ as the fluctuation induced light is associated with modes with $\omega \neq 0$. Substituting Eqs. (D3) and (D6) into Eq. (D2) we find that the bilateral spectrum is determined by:

$$\boldsymbol{\sigma}^b_{\text{ph},\omega,ij}(\Omega) = \frac{\mathcal{E}_\omega i}{(2\pi)^3} \int d^3\mathbf{k} \, \text{Tr}\left\{\hat{\mathbf{S}}_i \cdot \left[\mathbf{R}_{L\mathbf{k}} A_L(\omega) + \mathbf{R}^+_{T\mathbf{k}} A^+_{T\mathbf{k}}(\omega) + \mathbf{R}^+_{T\mathbf{k}} A^-_{T\mathbf{k}}(\omega)\right] \cdot \hat{\mathbf{S}}_j \cdot \right.$$
$$\left. \left[\mathbf{R}_{L\mathbf{k}} \delta(\omega - \omega_p) + \mathbf{R}_{L\mathbf{k}} \delta(\omega + \omega_p) + \mathbf{R}^+_{T\mathbf{k}} \delta(\omega - \omega_k) + \mathbf{R}^-_{T\mathbf{k}} \delta(\omega + \omega_k)\right]\right\}.$$

(D7)

In order to simplify further the formula, we take into account that $\text{Tr}\{\hat{\mathbf{S}}_i \cdot \mathbf{R}_{L\mathbf{k}} \cdot \hat{\mathbf{S}}_j \cdot \mathbf{R}_{L\mathbf{k}}\} = 0$, $\text{Tr}\{\hat{\mathbf{S}}_i \cdot \mathbf{R}^\pm_{T\mathbf{k}} \cdot \hat{\mathbf{S}}_j \cdot \mathbf{R}^\mp_{T\mathbf{k}}\} = 0$, and that:

$$\text{Tr}\{\hat{\mathbf{S}}_i \cdot \mathbf{R}_{L\mathbf{k}} \cdot \hat{\mathbf{S}}_j \cdot \mathbf{R}^\pm_{T\mathbf{k}}\} = \frac{c^4}{4\omega_k^2}(k^2 \delta_{ij} - k_i k_j) \tag{D8a}$$

$$\text{Tr}\{\hat{\mathbf{S}}_i \cdot \mathbf{R}^\pm_{T\mathbf{k}} \cdot \hat{\mathbf{S}}_j \cdot \mathbf{R}^\pm_{T\mathbf{k}}\} = \frac{2c^4}{\omega_k^2} k_i k_j. \tag{D8b}$$

where $\delta_{ij}$ is the Kronecker's symbol. Using these formulas, one finds that the bilateral spectrum tensor can be written as:

$$\boldsymbol{\sigma}^b_{\text{ph},\omega}(\Omega) = \frac{\mathcal{E}_\omega i}{(2\pi)^3} \int d^3\mathbf{k} \, \frac{c^4}{4\omega_k^2}\left(A_L(\omega)(k^2 \mathbf{1} - \mathbf{k} \otimes \mathbf{k}) + 8 A^+_{T\mathbf{k}}(\omega) \mathbf{k} \otimes \mathbf{k}\right) \delta(\omega - \omega_k) +$$
$$\frac{c^4}{4\omega_k^2}\left(A_L(\omega)(k^2 \mathbf{1} - \mathbf{k} \otimes \mathbf{k}) + 8 A^-_{T\mathbf{k}}(\omega) \mathbf{k} \otimes \mathbf{k}\right) \delta(\omega + \omega_k) +$$
$$\frac{c^4}{4\omega_k^2}\left(A^+_{T\mathbf{k}}(\omega)(k^2 \mathbf{1} - \mathbf{k} \otimes \mathbf{k}) + A^-_{T\mathbf{k}}(\omega)(k^2 \mathbf{1} - \mathbf{k} \otimes \mathbf{k})\right)\left[\delta(\omega - \omega_p) + \delta(\omega + \omega_p)\right]$$

(D9)

It is now convenient to switch to a system of spherical coordinates $k, \theta, \varphi$ in the wave vector space. The integrand depends on $\theta, \varphi$ only through the terms $\mathbf{k} \otimes \mathbf{k}$. Since $\iint d\theta d\varphi \sin\theta \, \mathbf{k} \otimes \mathbf{k} = \mathbf{1}_{3\times 3} 4\pi k^2 / 3$ it is clear that $\boldsymbol{\sigma}^b_{\text{ph},\omega}(\Omega)$ is a scalar determined by:

$$\sigma_{\text{ph},\omega}^{b}(\Omega) = \frac{\mathcal{E}_\omega i}{(2\pi)^3} 4\pi \int_0^\infty dk \; \frac{c^4 k^4}{4\omega_k^2}\left(\frac{2}{3} A_L(\omega) + \frac{8}{3} A_{Tk}^+(\omega)\right)\delta(\omega - \omega_k) +$$
$$\frac{c^4 k^4}{4\omega_k^2}\left(\frac{2}{3} A_L(\omega) + \frac{8}{3} A_{Tk}^-(\omega)\right)\delta(\omega + \omega_k) + \quad \text{(D10)}$$
$$\frac{c^4 k^4}{4\omega_k^2}\left(\frac{2}{3} A_{Tk}^+(\omega) + \frac{2}{3} A_{Tk}^-(\omega)\right)\left[\delta(\omega - \omega_p) + \delta(\omega + \omega_p)\right].$$

Feeding the above formula into (24a) with $A_L(\omega), A_{Tk}^+(\omega)$ evaluated with $\omega \to \omega + 0^+ i$ to take into account infinitesimal losses, one finds that the unilateral spectrum ($\omega > 0$) of the conductivity is given by:

$$\sigma_{\text{ph},\omega}(\Omega) = \frac{\mathcal{E}_\omega i}{12\pi^2} \int_0^\infty dk \; \frac{c^4 k^4}{\omega_k^2} \omega_p \left[\frac{1}{(\Omega - \Delta_k)\Delta_k} + \frac{1}{(\Omega + \Delta_k^*)\Delta_k^*} - \frac{1}{(\Omega - \tilde{\Delta}_k)\tilde{\Delta}_k} - \frac{1}{(\Omega + \tilde{\Delta}_k^*)\tilde{\Delta}_k^*}\right]\delta(\omega - \omega_k)$$
$$-\frac{c^4 k^4}{\omega_k^2} \omega_k \left[\frac{1}{(\Omega - \Delta_k)\Delta_k} + \frac{1}{(\Omega + \Delta_k^*)\Delta_k^*} + \frac{1}{(\Omega - \tilde{\Delta}_k)\tilde{\Delta}_k} + \frac{1}{(\Omega + \tilde{\Delta}_k^*)\tilde{\Delta}_k^*}\right]\delta(\omega - \omega_p)$$
(D11)

with $\Delta_k = +\omega_p - \omega_k - i0^+$ and $\tilde{\Delta}_k = -\omega_p - \omega_k - i0^+$. Note that the photonic conductivity has a spectrum with $\omega \geq \omega_p$, consistent with the spectrum of the bulk material.

As discussed in the main text, the real-part of the conductivity determines the irreversible light matter interactions. Evidently, it is determined by the contributions of the poles of the integrand of the form $\Omega - \Delta_k = 0$, $\Omega + \Delta_k^* = 0$, $\Omega - \tilde{\Delta}_k = 0$, $\Omega + \tilde{\Delta}_k^* = 0$. Supposing without loss of generality that $\Omega > 0$, only the poles of $\Omega + \Delta_k^* = 0$ and $\Omega + \tilde{\Delta}_k^* = 0$ contribute to the unilateral spectral density:

$$\text{Re}\{\sigma_{\text{ph},\omega}(\Omega)\} = \frac{\mathcal{E}_\omega}{12\pi} \int_0^\infty dk \; \frac{c^4 k^4}{\omega_k^2} \omega_p \left[\frac{1}{\Delta_k^0} \delta(\Omega + \Delta_k^0) - \frac{1}{\tilde{\Delta}_k^0} \delta(\Omega + \tilde{\Delta}_k^0)\right]\delta(\omega - \omega_k) +$$
$$-\frac{c^4 k^4}{\omega_k^2} \omega_k \left[\frac{1}{\Delta_k^0} \delta(\Omega + \Delta_k^0) + \frac{1}{\tilde{\Delta}_k^0} \delta(\Omega + \tilde{\Delta}_k^0)\right]\delta(\omega - \omega_p),$$
(D12)

with $\Delta_k^0 = +\omega_p - \omega_k < 0$ and $\tilde{\Delta}_k^0 = -\omega_p - \omega_k < 0$. Comparing the above formula with the modal expansion (C6), one can see that the dissipative terms emerge from interband transitions that start from the positive frequency transverse modes (terms in the first line) and positive frequency bulk plasmons (terms in the second line). All the transitions are between transverse modes and longitudinal modes. The delta-function $\delta(\Omega+\Delta_k^0)$ is associated with transitions between the positive frequency branches, whereas the delta-function $\delta(\Omega+\tilde{\Delta}_k^0)$ is associated with transitions between a positive frequency branch and a negative frequency branch.

The delta-function $\delta(\Omega+\Delta_k^0)$ is singular for the wave number $k_+ = \frac{1}{c}\sqrt{(\Omega+\omega_p)^2 - \omega_p^2}$, whereas the delta-function $\delta(\Omega+\tilde{\Delta}_k^0)$ is singular for $k_- = \frac{1}{c}\sqrt{(\Omega-\omega_p)^2 - \omega_p^2}$. The latter condition is only possible when $\Omega \geq 2\omega_p$. Taking into account that $\partial_k \omega_k = c^2 \frac{k}{\omega_k}$, one can show after some simplifications that the real-part of the conductivity satisfies:

$$\text{Re}\{\sigma_{\text{ph},\omega}(\Omega)\} = \frac{\mathcal{E}_\omega}{12\pi c}(\Omega+2\omega_p)\frac{\sqrt{(\Omega+\omega_p)^2 - \omega_p^2}}{\Omega+\omega_p}\left[-\omega_p \delta(\omega-(\Omega+\omega_p))+(\Omega+\omega_p)\delta(\omega-\omega_p)\right]$$
$$+\frac{\mathcal{E}_\omega}{12\pi c}(\Omega-2\omega_p)\frac{\sqrt{(\Omega-\omega_p)^2 - \omega_p^2}}{\Omega-\omega_p}\left[+\omega_p \delta(\omega-(\Omega-\omega_p))+(\Omega-\omega_p)\delta(\omega-\omega_p)\right]$$
(D13)

The term in the second line only contributes when $\Omega \geq 2\omega_p$, otherwise it should be ignored.

## Appendix E: Link between the photonic conductivity and the angular momentum

Here, we prove that in the weak dissipation limit, the spectral density of the photonic conductivity for a constant acceleration ($\Omega \to 0$) is strictly determined by the angular momentum of the thermal-light in the unperturbed cavity.

To show this, we note that in the lossless limit the (source-free) electrodynamics of the unperturbed cavity can be formulated as a Hermitian problem of the type $\hat{H}_g \mathbf{Q} = i \frac{d}{dt} \mathbf{Q}$, with $\hat{H}_g = \hat{H}_g(\mathbf{r}, -i\nabla)$ an Hermitian operator with respect to some weighted inner product $\langle ... | ... \rangle_g$ [65, 66]. Here, $\mathbf{Q} = \begin{pmatrix} \mathbf{f} & \mathbf{Q}^{(1)} & ... \end{pmatrix}^T$ is a state vector written in terms of the electromagnetic fields $\mathbf{f} = \begin{pmatrix} \mathbf{E} & \mathbf{H} \end{pmatrix}^T$ and of other internal degrees of freedom of the system ($\mathbf{Q}^{(\alpha)}$, $\alpha \geq 1$) due to the dispersive material response [65, 66]. In this framework, the electromagnetic modes of the system are the nontrivial solutions of the spectral problem $\hat{H}_g \mathbf{Q}_n = \omega_n \mathbf{Q}_n$.

Following Ref. [65], it is convenient to introduce the operator $\partial_i \hat{H}_g$ which is defined as $\partial_i \hat{H}_g = \frac{\partial}{\partial k_i} \hat{H}_g(\mathbf{r}, -i\nabla + \mathbf{k})$ evaluated for $\mathbf{k} = 0$. As demonstrated in [65], $\partial_i \hat{H}_g$ can be identified with the $i$-th component of the Poynting vector. In particular, for arbitrary state vectors $\mathbf{Q}_A = \begin{pmatrix} \mathbf{f}_A & \mathbf{Q}_A^{(1)} & ... \end{pmatrix}^T$ and $\mathbf{Q}_B = \begin{pmatrix} \mathbf{f}_B & \mathbf{Q}_B^{(1)} & ... \end{pmatrix}^T$ it is easy to prove that:

$$\langle \mathbf{f}_A | \hat{\mathbf{S}}_i | \mathbf{f}_B \rangle = \langle \mathbf{Q}_A | \partial_i \hat{H}_g | \mathbf{Q}_B \rangle_g. \tag{E1}$$

Here, $\hat{\mathbf{S}}_i$ is given by Eq. (A11) and $\langle ... | ... \rangle$ is the canonical inner product of the main text. The weighted inner product $\langle ... | ... \rangle_g$ is defined as in Ref. [65]. Thus, for a dispersive (local) material the operator $\partial_i \hat{H}_g$ acts only on the electromagnetic fields ($\mathbf{f}_A, \mathbf{f}_B$). Using the above result, one can rewrite the modal expansion (C7) of the photonic conductivity spectral density in terms of the eigenmodes of the operator $\hat{H}_g$:

$$\boldsymbol{\sigma}_{\text{ph},\omega,ij}\big|_{\Omega=0} = i \frac{\varepsilon_\omega}{V} \sum_{\substack{\omega_m \neq \omega_n, \\ \omega_n > 0}} \delta(\omega - \omega_n) \frac{\omega_m}{\omega_{mn}^2} \left[ -\langle \mathbf{Q}_n | \partial_i \hat{H}_g | \mathbf{Q}_m \rangle_g \langle \mathbf{Q}_m | \partial_j \hat{H}_g | \mathbf{Q}_n \rangle_g + \langle \mathbf{Q}_m | \partial_i \hat{H}_g | \mathbf{Q}_n \rangle_g \langle \mathbf{Q}_n | \partial_j \hat{H}_g | \mathbf{Q}_m \rangle_g \right]$$
(E2)

The normalization condition (C2) is equivalent to $\langle \mathbf{Q}_n | \mathbf{Q}_m \rangle_g = \delta_{n,m}$ [65, 66].

To proceed, we use the property (see Eq. (A6) of Ref. [65]):

$$\left\langle \mathbf{Q}_n | \partial_j \hat{H}_\mathrm{g} | \mathbf{Q}_m \right\rangle_\mathrm{g} = \frac{\omega_m - \omega_n}{i} \left\langle \mathbf{Q}_n | x_j | \mathbf{Q}_m \right\rangle_\mathrm{g}. \tag{E3}$$

This result holds true when the box is terminated with opaque-type (non-cyclic) boundary conditions, e.g., with PEC walls [65]. Thus, Eq. (E2) may be rewritten as:

$$\begin{aligned}
\boldsymbol{\sigma}_{\mathrm{ph},\omega,ij}\big|_{\Omega=0} &= i\frac{\mathcal{E}_\omega}{V} \sum_{\substack{\omega_m \neq \omega_n, \\ \omega_n > 0}} \delta(\omega-\omega_n)\omega_m \left[ -\left\langle \mathbf{Q}_n | x_i | \mathbf{Q}_m \right\rangle_\mathrm{g}\left\langle \mathbf{Q}_m | x_j | \mathbf{Q}_n \right\rangle_\mathrm{g} + \left\langle \mathbf{Q}_m | x_i | \mathbf{Q}_n \right\rangle_\mathrm{g}\left\langle \mathbf{Q}_n | x_j | \mathbf{Q}_m \right\rangle_\mathrm{g} \right] \\
&= i\frac{\mathcal{E}_\omega}{V} \sum_{\omega_n > 0} \delta(\omega-\omega_n)\left\langle \mathbf{Q}_n | x_j \hat{H}_\mathrm{g} x_i - x_i \hat{H}_\mathrm{g} x_j | \mathbf{Q}_n \right\rangle_\mathrm{g}
\end{aligned} \tag{E4}$$

In the last identity, we used the spectral theorem $\hat{H}_\mathrm{g} = \sum_{\omega_m} \omega_m |\mathbf{Q}_m\rangle\langle\mathbf{Q}_m|$. Taking into account that $\partial_i \hat{H}_\mathrm{g}$ can be expressed in terms of the commutator $[x_i, \hat{H}_\mathrm{g}] = x_i \hat{H}_\mathrm{g} - \hat{H}_\mathrm{g} x_i$ as $\partial_i \hat{H}_\mathrm{g} = -i[x_i, \hat{H}_\mathrm{g}]$ [65], one finds that $x_j \hat{H}_\mathrm{g} x_i - x_i \hat{H}_\mathrm{g} x_j = i(x_i \partial_j \hat{H}_\mathrm{g} - x_j \partial_i \hat{H}_\mathrm{g})$. Since $\partial_i \hat{H}_\mathrm{g}$ only acts on the electromagnetic degrees of freedom it can be identified with the Poynting vector operator ($\hat{\mathbf{S}}_i$). Thus, it follows that:

$$\boldsymbol{\sigma}_{\mathrm{ph},\omega,ij}\big|_{\Omega=0} = -\frac{\mathcal{E}_\omega}{V} \sum_{\omega_n > 0} \delta(\omega-\omega_n)\left\langle n | x_i \hat{S}_j - x_j \hat{S}_i | n \right\rangle. \tag{E5}$$

We switched back to the canonical inner product of the main text and to a state vector formed exclusively by the electromagnetic fields. The inner product $\langle n | x_i \hat{S}_j - x_j \hat{S}_i | n \rangle$ can be written explicitly in terms of the Abraham (kinetic) angular momentum of the $n$-th mode $\mathcal{L}_n = \frac{1}{c^2}\int_V dV\ \mathbf{r} \times \mathrm{Re}\{\mathbf{E}_n(\mathbf{r}) \times \mathbf{H}_n^*(\mathbf{r})\}$ as $\langle n | x_i \hat{S}_j - x_j \hat{S}_i | n \rangle = \varepsilon_{ijk} c^2 \mathcal{L}_{n,k}$ where $\varepsilon_{ijk}$ is the Levi-Civita symbol. Thus, the photonic spectral density satisfies:

$$\boldsymbol{\sigma}_{\mathrm{ph},\omega}\big|_{\Omega=0} = \frac{c^2}{V}\left[\mathcal{E}_\omega \sum_{\omega_n > 0} \delta(\omega-\omega_n)\mathcal{L}_n \right] \times \mathbf{1}. \tag{E6}$$

The term inside rectangular brackets is exactly the (thermal) fluctuation unilateral angular momentum spectral density $\mathcal{L}_\omega$ of the unperturbed cavity (see Eq. (20) of Ref. [65]) in the weak dissipation limit. Therefore, we have shown that:

$$\boldsymbol{\sigma}_{\text{ph},\omega}\Big|_{\substack{\Omega=0 \\ \text{weak dissipation}}} = c^2 \frac{1}{V} \mathcal{L}_\omega \times \mathbf{1}. \tag{E7}$$

It should be noted that from the fluctuation dissipation theorem [Eq. (14)] the (unilateral) angular momentum spectral density can also be expressed as:

$$\mathcal{L}_{\omega,k} = \mathcal{E}_\omega \frac{1}{\pi} \frac{1}{c^2} \operatorname{Re} \int d^3\mathbf{r} \, \operatorname{Tr}\left\{ \left( x_i \hat{\mathbf{S}}_j - x_j \hat{\mathbf{S}}_i \right) \cdot \frac{\overline{\mathcal{G}}(\mathbf{r},\mathbf{r},\omega)}{i\omega} \right\}, \tag{E8}$$

with $(i,j,k)$ identical to $(1,2,3)$ or to its positive permutations. As a side note, we mention that it is also possible to obtain Eq. (E7) through direct manipulations of Eq. (37) using Eq. (E8). The details are omitted here for conciseness.

It is straightforward to show that for reciprocal platforms the Green's function has the symmetry:

$$\overline{\mathcal{G}}(\mathbf{r},\mathbf{r}',\omega) = \boldsymbol{\sigma}_z \cdot \overline{\mathcal{G}}^T(\mathbf{r}',\mathbf{r},\omega) \cdot \boldsymbol{\sigma}_z, \qquad \text{with} \qquad \boldsymbol{\sigma}_z = \begin{pmatrix} \mathbf{1}_{3\times 3} & \mathbf{0}_{3\times 3} \\ \mathbf{0}_{3\times 3} & -\mathbf{1}_{3\times 3} \end{pmatrix}. \tag{E9}$$

Since $\hat{\mathbf{S}}_j$ is real-valued and transpose symmetric, we have that for reciprocal systems $\operatorname{Tr}\{\hat{\mathbf{S}}_j \cdot \overline{\mathcal{G}}(\mathbf{r},\mathbf{r},\omega)\} = \operatorname{Tr}\{\hat{\mathbf{S}}_j \cdot \overline{\mathcal{G}}^T(\mathbf{r},\mathbf{r},\omega)\} = \operatorname{Tr}\{\hat{\mathbf{S}}_j \cdot \boldsymbol{\sigma}_z \cdot \overline{\mathcal{G}}(\mathbf{r},\mathbf{r},\omega) \cdot \boldsymbol{\sigma}_z\}$. This also implies that $\operatorname{Tr}\{\hat{\mathbf{S}}_j \cdot \overline{\mathcal{G}}(\mathbf{r},\mathbf{r},\omega)\} = \operatorname{Tr}\{\boldsymbol{\sigma}_z \cdot \hat{\mathbf{S}}_j \cdot \boldsymbol{\sigma}_z \cdot \overline{\mathcal{G}}(\mathbf{r},\mathbf{r},\omega)\}$. But explicit calculations, using Eq. (A11), show that $\boldsymbol{\sigma}_z \cdot \hat{\mathbf{S}}_j \cdot \boldsymbol{\sigma}_z = -\hat{\mathbf{S}}_j$. This proves that for reciprocal platforms $\operatorname{Tr}\{\hat{\mathbf{S}}_j \cdot \overline{\mathcal{G}}(\mathbf{r},\mathbf{r},\omega)\} = 0$, and thereby the fluctuation induced angular momentum vanishes [57, 63]:

$$\mathcal{L}_\omega\Big|_{\substack{\text{reciprocal}\\ \text{systems}}} = 0. \tag{E10}$$

Note that the above formula holds true even in case of strong loss. Evidently, it implies that for reciprocal platforms the constant-acceleration photonic conductivity vanishes in the weak dissipation limit: $\left.\boldsymbol{\sigma}_{\text{ph},\omega}\right|_{\substack{\Omega=0 \\ \text{weak dissipation} \\ \text{reciprocal system}}} = 0$.

# References

[1] F. D. M. Haldane, S. Raghu, "Possible realization of directional optical waveguides in photonic crystals with broken time-reversal symmetry", *Phys. Rev. Lett.*, **100**, 013904, (2008).

[2] S. Raghu, F. D. M. Haldane, "Analogs of quantum-Hall-effect edge states in photonic crystals", *Phys. Rev. A*, **78**, 033834, (2008).

[3] T. Ozawa, H. M. Price, A. Amo, N. Goldman, M. Hafezi, L. Lu, M. C. Rechtsman, D. Schuster, J. Simon, O. Zilberberg, I. Carusotto, "Topological Photonics", *Rev. Mod. Phys.* **91**, 015006, (2019).

[4] L. Lu, J. D. Joannopoulos, M. Soljačić, "Topological photonics", *Nat. Photonics*, **8**, 821, (2014).

[5] Z. Wang, Y. Chong, J. D. Joannopoulos, and M. Soljacic, "Observation of unidirectional backscattering immune topological electromagnetic states," *Nature* **461**, 772–775, **(**2009).

[6] A. B. Khanikaev, S H. Mousavi, W.-K. Tse, M. Kargarian, A. H MacDonald, and G. Shvets, "Photonic topological insulators," *Nat. Materials* 12, 233, (2013).

[7] M. Hafezi, E. A. Demler, M. D. Lukin, J. M. Taylor, "Robust optical delay lines with topological protection", *Nat. Phys.*, **7**, 907 (2011).

[8] M. G. Silveirinha, "Chern Invariants for Continuous Media", *Phys. Rev. B*, **92**, 125153, (2015).

[9] S. A. H. Gangaraj, M. G. Silveirinha, G. W. Hanson, "Berry phase, Berry Connection, and Chern Number for a Continuum Bianisotropic Material from a Classical Electromagnetics Perspective", *IEEE J. Multiscale and Multiphys. Comput. Techn.*, **2**, 3, (2017).

[10] R. B. Laughlin, "Quantized Hall Conductivity in Two Dimensions" *Phys. Rev. B*, **23**, 5632(R) (1981).

[11] D. J. Thouless, M. Kohmoto, M. P. Nightingale, and M. den Nijs, "Quantized Hall Conductance in a Two-Dimensional Periodic Potential", *Phys. Rev. Lett.*, **49**, 405, (1982).

[12] B. I. Halperin, "Quantized Hall conductance, current-carrying edge states, and the existence of extended states in a two-dimensional disordered potential", *Phys. Rev. B*, **25**, 2185, (1982).

[13] Y. Hatsugai, "Chern number and edge states in the integer quantum Hall effect", *Phys. Rev. Lett.* **71**, 3697, (1993).

[14] G. Barton, C. Eberlein, "On quantum radiation from a moving body with finite refractive index" *Ann. Phys.*, **227**, 222, (1993).

[15] G. Barton, "The quantum radiation from mirrors moving sideways" *Ann. Phys.*, **245**, 361, (1996).

[16] J. B. Pendry, "Quantum Friction—Fact or Fiction?", *New J. Phys.* **12**, 033028 (2010).

[17] S. I. Maslovski, M. G. Silveirinha, "Quantum friction on monoatomic layers and its classical analog", *Phys. Rev. B*, **88**, 035427, (2013).

[18] M. G. Silveirinha, "Quantization of the Electromagnetic Field in Non-dispersive Polarizable Moving Media above the Cherenkov Threshold", *Phys. Rev. A*, **88**, 043846, (2013).

[19] M. G. Silveirinha, "Theory of Quantum Friction", *New J. Phys.*, **16**, 063011 (1-29), (2014).

[20] M. G. Silveirinha, "Optical instabilities and spontaneous light emission by polarizable moving matter", *Phys. Rev. X*, **4**, 031013, (2014).

[21] V. Macrì, A. Ridolfo, O. Di Stefano, A. F. Kockum, F. Nori and S. Savasta, "Nonperturbative Dynamical Casimir Effect in Optomechanical Systems: Vacuum Casimir-Rabi Splittings", *Phys. Rev. X* **8**, 011031 (2018).

[22] D. Oue and M. Matsuo, "Twisting an optomechanical cavity", *Phys. Rev. A* **106**, L041501 (2022).

[23] V. V. Dodonov, "Current status of the dynamical Casimir effect", *Phys. Scr.* **82**, 038105, (2010).

[24] V. Dodonov, "Fifty Years of the Dynamical Casimir Effect", *Physics,* **2**, 67 (2020).

[25] W.G. Unruh, "Notes on black-hole evaporation", *Phys. Rev. D* **14**, 870 (1976).

[26] G. T. Moore, "Quantum theory of the electromagnetic field in a variable-length one-dimensional cavity", *J. Math. Phys.* **11**, 2679 (1970).

[27] E. Yablonovitch, "Accelerating reference frame for electromagnetic waves in a rapidly growing plasma: Unruh-Davies-Fulling-Dewitt radiation and the nonadiabatic Casimir effect", *Phys. Rev. Lett.* **62**, 1742 (1989).

[28] J. R. Johansson, G. Johansson, C. M. Wilson, F. Nori, "Dynamical Casimir effect in a superconducting coplanar waveguide", *Phys. Rev. Lett.* **103**, 147003 (2009).

[29] C. M. Wilson, T. Duty, M. Sandberg, F. Persson, V. Shumeiko, and P. Delsing, "Photon generation in an electromagnetic cavity with a time-dependent boundary", *Phys. Rev. Lett.* **105**, 233907 (2010).

[30] C. M. Wilson, G. J. Johansson, A. Pourkabirian, M. Simoen, J. R. Johansson, T. Duty, F. Nori, and P. Delsing, "Observation of the dynamical Casimir effect in a superconducting circuit", *Nature* **479**, 376 (2011).

[31] C. Eberlein, "Sonoluminescence as quantum vacuum radiation", *Phys. Rev. Lett.*, **76**, 3842, (1996).

[32] T. Ozawa and I. Carusotto, "Anomalous and Quantum Hall Effects in Lossy Photonic Lattices", *Phys. Rev. Lett.*, **112**, 133902 (2014).

[33] M. Tarnowski F. Nurunal, N. Fläschner, B. S. Rem, A. Eckardt, K. Sengstock, and C. Weitenberg, "Measuring topology from dynamics by obtaining the Chern number from a linking number", *Nat. Comm.* **10**, 1728, (2019).

[34] E. Galiffi, R. Tirole, S. Yin, H. Li, S. Vezzoli, P. A. Huidrobo, M. G. Silveirinha, R. Sapienza, A. Alù and J. B. Pendry, "Photonics of time-varying media," *Adv. Photon.* **4**, 014002 (2022).

[35] Z. Deck-Léger, N. Chamanara, M. Skorobogatiy, M. G. Silveirinha and C. Caloz, "Uniform-velocity spacetime crystals," *Adv. Photon.* **1**, 1 – 26 (2019).

[36] Y. Mazor and A. Alù, "One-way hyperbolic metasurfaces based on synthetic motion", *IEEE Trans. Antennas Propag.*, **68**, 1739 (2020).

[37] M. Kreiczer and Y. Hadad, "Wave analysis and homogenization of a spatiotemporally modulated wire medium", *Phys. Rev. Applied*, **16**, 054003 (2021).

[38] H. T. To, "Homogenization of dynamic laminates", *J. Math. Anal. Appl.* **354**, 518, (2009).

[39] P. A. Huidobro, E. Galiffi, S. Guenneau, R. V. Craster and J. B. Pendry, "Fresnel drag in space–time-modulated metamaterials," *Proc. Natl. Acad. Sci. USA* 116, 24943 (2019).

[40] P. Huidobro, M. G. Silveirinha, E. Galiffi and J. B. P. Pendry, "Homogenization Theory of Space-Time Metamaterials," *Phys. Rev. Applied* **16**, 014044, (2021).

[41] F. R. Prudêncio and M. G. Silveirinha, "Synthetic Axion Response with Spacetime Crystals," *Phys. Rev. Applied* 19, 024031, (2023).

[42] D. Oue, K. Ding, J. B. Pendry, "Noncontact frictional force between surfaces by peristaltic permittivity modulation", *Phys. Rev. A* **107**, 063501, (2023).

[43] S. A. R. Horsley, J. B. Pendry, "Time varying gratings model Hawking radiation", arXiv:2302.04066.

[44] E. Galiffi, P. A. Huidrobo and J. B. Pendry, "An Archimedes' screw for light," *Nat. Commun.* 13, 2523 (2022).

[45] A. Alex-Amor, C. Molero, M. G. Silveirinha, "Analysis of Metallic Spacetime Gratings using Lorentz Transformations", *Phys. Rev. Appl.*, **20**, 014063, (2023)..

[46] F. R. Prudêncio, M. G. Silveirinha, "Replicating Physical Motion with Minkowskian Isorefractive Spacetime Crystals", *Nanophotonics*, **10**.1515/nanoph-2023-0144,( 2023).

[47] D. L. Sounas, C. Caloz, A. Alù, "Giant non-reciprocity at the subwavelength scale using angular momentum-biased metamaterials", *Nat. Comm.*, **4** (1), 1-7 (2013).

[48] N.A. Estep, D.L. Sounas, J. Soric, A. Alù, "Magnetic-free non-reciprocity and isolation based on parametrically modulated coupled-resonator loops", *Nat. Phys.* **10**, 923 (2014).

[49] D. L. Sounas, A. Alù, "Angular-momentum-biased nanorings to realize magnetic-free integrated optical isolation", *ACS Photonics* **1**, 198 (2014).

[50] J. C. Serra, M. G. Silveirinha, "Rotating Spacetime Modulation: Topological Phases and Spacetime Haldane Model", *Phys. Rev. B*, **107**, 035133, (2023).

[51] R. N. C. Pfeifer, T. A. Nieminen, N. R. Heckenberg, and H. R.-Dunlop, "Colloquium: Momentum of an electromagnetic wave in dielectric media", *Rev. Mod. Phys.*, **79**, 1197 (2007).

[52] N. L. Balazs, "The Energy-Momentum Tensor of the Electromagnetic Field inside Matter", *Phys. Rev.* **91**, 408 (1953).

[53] S. M. Barnett, "Resolution of the Abraham-Minkowski Dilemma", *Phys. Rev. Lett.*, **104**, 070401, (2010).

[54] M. G. Silveirinha, "Reexamination of the Abraham-Minkowski Dilemma", *Phys. Rev. A*, **96**, 033831, (2017).

[55] J. D. Jackson, Classical Electrodynamics (Wiley, New York, 1998).

[56] L. D. Landau, E. M. Lifshitz, Statistical Physics Part 2, vol. 9 of Course on Theoretical Physics, Pergamon Press, Oxford, 1981 (Chapter VIII, p. 320).

[57] M. G. Silveirinha, "Topological Angular Momentum and Radiative Heat Transport in Closed Orbits", Phys. Rev. B, **95**, 115103, (2017).

[58] L. Onsager "Reciprocal Relations in Irreversible Processes. I.", *Phys. Rev.* **37** 405, (1931).

[59] F. Rohrlich, "The self-force and radiation reaction", *Am. J. Phys.*, **68**, 1109, (2000).

[60] J. N. Winn, S. Fan, J. D. Joannopoulos, and E. P. Ippen, "Interband transitions in photonic crystals", *Phys. Rev. B* **59**, 1551 (1999).

[61] Z. Yu and S. Fan, "Complete optical isolation created by indirect interband photonic transitions," *Nat. Photon.* **3**, 91 (2009).

[62] S. A. R. Horsley and S. Bugler-Lamb, "Negative frequencies in wave propagation: A microscopic model", *Phys. Rev. A* **93**, 063828, (2016).

[63] L. Zhu, S. Fan, "Persistent Directional Current at Equilibrium in Nonreciprocal Many-Body Near Field Electromagnetic Heat Transfer", *Phys. Rev. Lett.*, **117**, 134303 (2016).

[64] M. G. Silveirinha, "Quantized Angular Momentum in Topological Optical Systems", *Nat. Comm.*, **10**, 349, (2019).

[65] M. G. Silveirinha, "Proof of the bulk-edge correspondence through a link between topological photonics and fluctuation-electrodynamics", *Phys. Rev. X*, **9**, 011037 (2019).

[66] M. G. Silveirinha, "Modal expansions in dispersive material systems with application to quantum optics and topological photonics", chapter in "Advances in Mathematical Methods for Electromagnetics", (edited by Paul Smith, Kazuya Kobayashi) published by the IET, 2021.

[67] M. G. Silveirinha, "Topological Classification of Chern-type insulators by means of the Photonic Green Function", *Phys. Rev. B*, **97**, 115146, (2018).

[68] P. Ben-Abdallah, "Photon Thermal Hall Effect", *Phys. Rev. Lett.* **116**, 084301 (2016).